\documentclass[12pt]{article}
\newlength{\abstractwidth}
\renewcommand{\thefootnote}{\fnsymbol{footnote}}
\renewcommand{\thanks}[1]{\footnote{#1}} 
\newcommand{\starttext}{
\setcounter{footnote}{0}
\renewcommand{\thefootnote}{\arabic{footnote}}}

\newcommand{\bea}{\begin{eqnarray}}
\newcommand{\eea}{\end{eqnarray}}
\newcommand{\<}{\langle}
\renewcommand{\>}{\rangle}

\def\A{{\cal A}}
\def\B{{\cal B}}
\def\C{{\cal C}}

\def\F{{\cal F}}

\def\K{{\cal K}}
\def\L{{\cal L}}

\def\O{{\cal O}}

\def\Q{{\cal Q}}
\def\R{{\cal R}}
\def\U{{\cal U}}
\def\V{{\cal V}}

\def\Y{{\cal Y}}

\def\B{{\cal B}}

\def\det{{\rm det}}
\def\sdet{{\rm sdet}}

\def\half{ {1\over 2}}
\def\z{{\bf z}}

\def\p{\partial}
\def\pz{\partial _z}
\def\pu{\partial _u}

\def\pw{\partial _w}
\def\tet{\vartheta}
\def\chiz{{\chi _{\bar z}{} ^+}}
\def\chiw{{\chi _{\bar w}{} ^+}}
\def\chiu{{\chi _{\bar u}{} ^+}}
\def\chiv{{\chi _{\bar v}{} ^+}}

\def\e{\epsilon}

\def\no{\nonumber}

\begin{document}
\starttext
\baselineskip=16pt
\setcounter{footnote}{0}

\begin{flushright}
UCLA/05/TEP/2 \\
Columbia/Math/05 \\
2005 January 24 \\
\end{flushright}

\bigskip

\begin{center}
{\Large\bf TWO-LOOP SUPERSTRINGS V}\\
\bigskip 
{\large \bf Gauge Slice Independence of the N-Point Function}
\footnote{Research supported in part by National Science
Foundation grants PHY-01-40151 and DMS-02-45371.}

\bigskip\bigskip
{\large Eric D'Hoker$^a$ and D.H. Phong$^b$}

\bigskip
$^a$ \sl Department of Physics and Astronomy \\
\sl University of California, Los Angeles, CA 90095, USA\\
$^b$ \sl Department of Mathematics\\
\sl Columbia University, New York, NY 10027, USA

\end{center}

\bigskip\bigskip

\begin{abstract}

A systematic construction of superstring scattering amplitudes for $N$ 
massless NS bosons to two loop order is given, based on the projection 
of supermoduli space onto super period matrices used earlier for the 
superstring measure in the first four papers of this series. 
The one important new difficulty arising for the $N$-point amplitudes is 
the fact that the projection onto super period matrices introduces 
corrections to the chiral vertex operators for massless NS bosons 
which are not pure (1,0) differential forms. 
However, it is proved that the chiral amplitudes are closed differential forms, 
and transform by exact differentials on the worldsheet
under changes of gauge slices. 
Holomorphic amplitudes and independence of left from right movers
are recaptured after the extraction of terms which are Dolbeault exact in 
one insertion point, and de~Rham closed in the remaining points. 
This allows a construction of GSO projected, integrated superstring 
scattering amplitudes which are independent of the choice of gauge slices
and have only physical kinematical singularities. 

\end{abstract}

\vfill\eject

\baselineskip=15pt
\setcounter{equation}{0}
\setcounter{footnote}{0}

\section{Introduction}

In the papers \cite{I,II,III,IV} of this series on superstring 
perturbation theory at two-loops, a gauge slice independent chiral 
superstring measure has been derived from first principles. 
The key procedure is a projection of supergeometries onto their 
super period matrices instead of onto the standard period matrices. 
This projection led to a new stress tensor correction as well as
to new finite-dimensional determinant factors, both of which together 
rid the measure of all the ambiguities plaguing it in the past.

\medskip

For the $N$-point function, it suffices to apply the same gauge-fixing 
procedure, after insertion of appropriate vertex operators. 
The one important subtlety is the emergence of some new corrections
to the vertex operators, due to the deformation of complex
structures inherent to the projection onto super period matrices.

\medskip

The main goal of the present paper is to establish the gauge slice 
independence of the resulting $N$-point scattering amplitudes, 
when the corrections  to the vertex operators are taken properly into account.  
Complete gauge slice independence will guarantee the absence of unphysical kinematical singularities in the full scattering amplitudes.

\medskip

Underlying the gauge slice independence of the scattering
amplitudes are two key results on the chiral amplitudes (or chiral blocks). 
The first is the fact that the chiral  amplitudes are closed differential forms 
in the vertex operator insertion points, and transform by the addition of exact differentials on the worldsheet under change of slice. 
The second is a remarkable relation between 
superholomorphicity and holomorphicity: superholomorphic correlation functions,
which are forms on products of super Riemann surfaces, descend to holomorphic forms on the products of the corresponding Riemann surfaces,
modulo forms which are Dolbeault $\bar\p$-exact in one variable
and de~Rham closed in the remaining variables. 
As a by-product, we obtain many new holomorphic sections of vector 
bundles over the moduli space of Riemann surfaces. These sections may be mathematically interesting in their own right.

\medskip

We describe now our results in greater detail.

\subsection{Corrections to the Vertex Operators}

The key new subtlety in the evaluation of the $N$-point function
originates from the fact that, in accordance with the projection onto 
the super period matrices $\hat\Omega_{IJ}$, all vertex operators 
must be deformed to the complex structure defined by $\hat\Omega_{IJ}$. 
Now the deformation of the correlation functions is accounted for by the 
same stress tensor insertion as in \cite{I,II,III,IV} (see also the review \cite{dp02}). 
However, a covariant 
amplitude must incorporate the volume form $d^{2|2}{\bf z}\,E({\bf z})$ 
at the vertex position ${\bf z}$
on the super Riemann surface, and this volume form has to be deformed as well. Incorporating this volume form and its deformation, and integrating over 
the $\theta$ super-coordinate of ${\bf z}=(z,\theta)$, we obtain the following  
component expression  for the full chiral vertex  of NS massless bosons,
\bea
\label{vertexsum}
\V(z; \epsilon, k)
=
\V^{(0)}(z; \epsilon, k)+\V^{(1)}(z;\epsilon, k)+\V^{(2)}(z;\epsilon, k)
\eea
where 
\bea
\label{vertexdecomp}
\V^{(0)}(z; \epsilon, k) 
& = & 
\epsilon^\mu\,dz(\p_zx_+^\mu-ik^\nu\psi_+^\mu\psi_+^\nu)(z)\,
e^{ik \cdot x_+ (z)}
\no \\
\V^{(1)}(z; \epsilon, k)
&=&
-\half \e^\mu d\bar z\chiz\psi_+^\mu(z) \, e^{ik \cdot x_+ (z)}
\no \\
\V^{(2)}(z; \epsilon, k)
&=&
- \epsilon^ \mu\hat\mu_{\bar z}{}^zd\bar z(\p_zx_+^\mu
-ik^\nu\psi_+^\mu\psi_+^\nu)(z)\,e^{ik \cdot x_+ (z)}
\eea
Here, $\V ^{(0)} (z; \epsilon, k)$ is the familiar vertex operator ($x_+(z)$ and 
$\psi _+(z)$ denote respectively the effective chiral scalar boson and the 
chiral fermion),  and $\V^{(1)}(z; \epsilon, k)$ 
and $\V^{(2)}(z; \epsilon, k)$ are corrections which 
depend both on the gravitino slice $\chiz$ and the Beltrami differential 
$\hat\mu_{\bar z}{}^z$ for the passage from
period matrix to super period matrix. In particular, the full vertex operators 
are gauge slice dependent.

\medskip

We stress that $\V^{(0)}(z; \epsilon, k)$ is a $(1,0)$ form, but 
$\V^{(1)}(z; \epsilon, k)$ and $\V^{(2)}(z; \epsilon, k)$ are $(0,1)$ 
forms, so that the full vertex operator $\V(z; \epsilon, k)$ includes 
both $(1,0)$ and $(0,1)$ components. Although $\V^{(1)}(z; \epsilon, k)$ 
and particularly $\V^{(2)}(z; \epsilon, k)$ are the source of many complications, 
their omission would certainly lead to unacceptable gauge-dependent 
results for the final superstring amplitudes.

\subsection{The Chiral Amplitudes ${\cal B}[\delta]$}

Let ${\cal B}[\delta]$ denote the full chiral amplitude for the scattering of $N$ 
massless NS bosons, incorporating the chiral measure, ghost, and superghost 
contributions which were derived in \cite{I,II}. 
The explicit expression for $\B [\delta]$ is given 
by\footnote{We use the new notation $\B [\delta]$, in order to distinguish
this quantity from the chiral block $\F [\delta]$ derived in \cite{dp89} and 
to be described in (\ref{block1}). The difference between the two is that 
the definition of $\F [\delta]$ does not include the effects of the measure factors 
$d^{2|2} {\bf z_i}  E({\bf z}_i)$ at the vertex operators, while $\B [\delta]$ does.
Also, the integration over the worldsheet $\Sigma$ will be abbreviated by 
$\int _\Sigma \to \int$, and the vertex operators by 
$\V_i = \V _i (z_i,\epsilon_i , k_i)$ when no confusion is 
expected to arise.}
\bea
\label{calb}
{\cal B}[\delta] (z_i; \epsilon_i, k_i,p_I)
&=&\prod_{I\leq J}d\Omega_{IJ}\,\int\,\prod_{\alpha=1,2} d\zeta^\alpha
\ {\prod_{a=1}^3b(p_a)\prod_{\alpha=1}^2\delta(\beta(q_{\alpha}))
\over
{\rm det}\,\Phi_{IJ+}(p_a)\,{\rm det}\,\<H_{\alpha}|\Phi_\beta^*\>}
\\
&&
\hskip .6in 
\times
\bigg \< Q(p_I)\,
\exp \bigg  \{ {1\over 2\pi}\int \big (\chi S +\hat\mu T \big ) \bigg \} \,
\prod_{j=1}^N\V_j  \bigg \>
\no
\eea
The notations used here are those of \cite{II}: 
the integration is over the odd super-moduli $\zeta ^\alpha, ~ \alpha =1,2 $;
$S$ and $T$ are the supercurrent and stress tensor respectively; 
$\Phi _{IJ+}$ and $\Phi ^* _\alpha$ are super-holomorphic 3/2 differentials;
and $H_\alpha$  dual super-Beltrami differentials;
$p_I$ are the internal loop momenta required by chiral splitting;
and $ Q(p_I) = \exp \{ i p_I ^\mu \oint _{B_I} dz \p_z x_+^\mu(z) \}$.

\medskip

The amplitude $\B [\delta]$ may be decomposed into the sum of a connected 
part $\B [\delta]  ^{(c)}$, and a disconnected part $\B [\delta ]^{(d)}$,
\bea
\label{calB}
\B [\delta]  \hskip .15in & = & \B [\delta ] ^{(d)} + \B [\delta ] ^{(c)} 
\no \\
\B [\delta ] ^{(d)} & = &
d \mu _2 [\delta]   ~ \left \< Q(p_I) 
\prod _{i=1}^N \V _i ^{(0)} (z_i,\epsilon_i,k_i) \right \>
\no \\
\B [\delta ] ^{(c)}  &= & d \mu _0 [\delta ] 
\int \prod _{\alpha =1,2} d \zeta ^\alpha
\bigg ( \Y_1 + \Y_2 + \Y_3 + \Y_4 + \Y_5 \bigg )
\eea
The {\sl disconnected part} consists of  the 
finite-dimensional  determinants, and the self-con-tractions of 
the stress tensor $T$ and  the supercurrents $SS$. It was  
already evaluated in \cite{I}. 
The {\sl connected part} is defined to be $\B [\delta] - \B [\delta ] ^{(d)} $.
The prefactors  $d \mu _0 [\delta]$ and $d \mu _2 [\delta]$
are components of the chiral measure $\A [\delta]$, 
defined and evaluated in \cite{II},
\bea
\label{measure}
\A [\delta]  \prod_{I\leq J}d\hat\Omega_{IJ}
&=& 
d\mu _0 [\delta] + \zeta ^1 \zeta ^2 d\mu _2[\delta]
\no \\ && \no \\
d\mu _0 [\delta]
&=& 
{\< \prod _{a=1} ^3 b(p_a) \prod _{\alpha =1}^2 \delta (\beta (q_\alpha))
\over \det \omega _I \omega _J (p_a)} \prod_{I\leq J}d\hat\Omega_{IJ}
\no \\ && \no \\
d\mu _2 [\delta ]
&=&
{ \tet [\delta ](0,\hat\Omega)^4 \Xi _6 [\delta](\hat\Omega)
\over 16 \pi ^6 ~ \Psi _{10}(\hat\Omega)} \prod_{I\leq J}d\hat\Omega_{IJ}
\eea
The quantities $\A [\delta] $, $d\mu _0 [\delta] $ and $d\mu_2[\delta] $ 
are independent of the ghost insertion points $p_a$. 
While $d\mu_2[\delta] $ is also independent of $q_1$ and $q_2$, 
$d\mu _0 [\delta] $ is not, but this dependence should disappear from 
the full amplitude which combines also the $\zeta$-dependent matter part. 
The various  components $\Y _i, ~ i=1,\cdots, 5$ are given by the following correlators,
\bea
\label{Ys}
\Y _1 
& = & 
{1 \over 8 \pi ^2} \left \< Q(p_I)\,  \int \! \chi S ~ \int \! \chi S ~
\prod _{i=1}^N \V_i ^{(0)} \right \> _{(c)}
\no \\
\Y _2 
& = & 
{1 \over 2 \pi } \left \< Q(p_I)\, \int \! \hat \mu T ~
\prod _{i=1}^N \V_i ^{(0)} \right \> _{(c)}
\no \\
\Y _3 
& = & 
{1 \over 2 \pi } \sum _{i=1} ^N \left \< Q(p_I)\, \int \! \chi S ~ \V ^{(1)} _i
~ \prod _{j \not= i}^N  \V_j ^{(0)}  \right \> 
\no \\
\Y _4 
& = & 
\half \sum _{i \not= j} \left \< Q(p_I) ~ 
 \V^{(1)} _i  ~ \V^{(1)} _j  ~ \prod _{l \not= i,j}^{N}  \V_l ^{(0)} \right \> 
\no \\
\Y _5 
& = & 
\sum _{i =1}^N \left \< Q(p_I) ~  \V ^{(2)} _i ~ 
\prod _{j \not= i}^{N}  \V_j ^{(0)} \right \> 
\eea
The polarization vectors $\epsilon _i$ will be viewed as anti-commuting 
with themselves and with the differential 1-forms $dz_i$ and $d\bar z_i$,  
so that the vertices $\V_i$  commute with one another, and their ordering 
in the above correlators is arbitrary. This formal device was introduced and 
used extensively in \cite{dp89,dp88} and will be useful also here.
Since all dependence on $\epsilon _i$  in the amplitudes  is linear, this 
prescription reproduces the amplitudes. Finally, the $\delta$-dependence 
of $\Y_1 , \cdots , \Y_5$ will be understood throughout.

\medskip

All quantities above are expressed with respect to the  superperiods 
$\hat \Omega _{IJ}$. We recall the relations between the period 
matrix $\Omega _{IJ}$, the super-period matrix $\hat \Omega _{IJ}$
and the Beltrami differential $\hat \mu _{\bar z} {}^z$
which provides the complex structure
deformation between $\Omega _{IJ}$ and $\hat \Omega _{IJ}$,
\bea
\label{OmegaDef}
\Omega _{IJ} - \hat \Omega _{IJ} 
& = &
{ i \over 8 \pi } \int \! d^2z \int \! d^2w \omega _I(z) \chiz 
S_\delta (z,w) \chiw \omega _J (w)
\no \\
& = & 
i \int \! d^2 z \hat \mu _{\bar z} {} ^z  \omega _I (z) \omega _J (z)
\eea
The vacuum expectation values in $\Y_1$ and $\Y_2$ 
are connected, as indicated by the subscript $\< \cdots \> _{(c)}$, in the following sense.
All self-contractions of $T$ are to be excluded. The contributions
in which both $x_+$ and $\psi _+$ are contracted between the two 
supercurrents $SS$ are to be excluded. The contractions of only 
a single field between the two $S$-operators, with the remaining operators contracted elsewhere, however, are to be included in $\Y_2$. 

\medskip

The chiral blocks ${\cal B}[\delta]$ are 1-forms (including both $(1,0)$ 
and $(0,1)$ components) in 
each vertex point $z_i$ with the following monodromy,
\bea
\label{monodromy}
{\cal B}[\delta] (z_i+\delta_{ij} A_K,\theta_i; \e_i,k_i,p_I)
& = &
{\cal B}[\delta] (z_i,\theta_i; \e_i,k_i,p_I)
\no \\
{\cal B}[\delta] (z_i+\delta_{ij}B_K,\theta_i; \e_i, k_i,p_I)
& = &
{\cal B}[\delta] (z_i,\theta_i; \e_i,k_i,p_I+\delta_{IK}k_j).
\eea
Thus they should be viewed as sections of a flat vector bundle
over the moduli space of Riemann surfaces with $N$-punctures.

\subsection{The Structure of Chiral and Holomorphic Blocks}

\medskip

It is a fundamental principle of string theory, reflecting the independence 
of left and right-movers, that the superstring scattering amplitudes be 
realizable as hermitian pairings of  {\sl holomorphic} sections of vector 
bundles over the moduli space of Riemann surfaces with $N$ punctures.
When $N>0$, an old puzzle of superstring perturbation theory is how to 
reconcile this holomorphicity requirement with the fact that its building blocks, 
namely the scalar superfield chiral  amplitudes ${\cal F}[\delta]$, are only 
{\sl superholomorphic}. For non-zero gravitino field $\chiz$,
the notion of superholomorphicity appears fundamentally different from 
the notion of holomorphicity.

\medskip

The vertex operator corrections $\V^{(1)}$ and $\V^{(2)}$
seemingly compound the problem: they are $(0,1)$-forms,
and such forms do not even admit an intrinsic notion of holomorphicity, 
since their covariant derivative $\nabla_{\bar z}$ requires a connection.

\medskip

Perhaps surprisingly, exactly the opposite is the case. The $(0,1)$ 
vertex operator correction terms provide precisely the
mechanism for restoring holomorphicity to the string amplitudes,
and this is in fact closely related to the ultimate gauge slice independence 
of the $N$-point function: with the $(0,1)$
correction terms, all the terms in the chiral amplitude ${\cal B}[\delta]$ 
which are not already tensor products of $(1,0)$-forms in all insertion points,
combine into a sum $\sum_{i=1}^N\bar\p_i{\cal S}_i[\delta](z;\epsilon,k,p_I)$ 
of terms which are Dolbeault
$\bar\p$-exact in one variable, and de~Rham $d$-closed in the
other variables. Upon completion of the Dolbeault $\bar\p$-exact 
differential into a de Rham $d$-exact  differential,
\bea
d_i{\cal S}_i[\delta](z;\epsilon,k,p_I)
=
\bar\p_i{\cal S}_i[\delta](z;\epsilon,k,p_I)
+
\p_i{\cal S}_i[\delta](z;\epsilon,k,p_I),
\eea
the de~Rham $d$-exact terms $ d_i {\cal S} _i [\delta]$ drop out of any 
physical amplitude, and the chiral amplitude ${\cal B}[\delta]$ is 
effectively replaced by the amplitude
${\cal H}[\delta]={\cal B}[\delta]-
\sum_{i=1}^N\p_i{\cal S}_i[\delta](z;\epsilon,k,p_I)$,
which is both holomorphic and gauge slice independent.

\medskip

The precise statements are the following:

\bigskip
 
\noindent {\bf \large (a) Closedness}
\smallskip

The forms ${\cal B}[\delta]$ are closed in each variable $z_j$;

\medskip \smallskip

\noindent {\bf \large (b) Slice-change by exact differentials on the worldsheet}
\smallskip

Under infinitesimal changes of either the gravitino slice $\chi$ or the 
Beltrami differential $\hat\mu$, the forms ${\cal B}[\delta]$ change by 
terms which are de~Rham $d$-exact in one variable and de~Rham
$d$-closed in all other variables.\footnote{Henceforth,  we shall  denote 
the dependence of the blocks on $(z_i; \e_i, k_i, p_I)$ simply by $(z; \e,k,p_I)$.}
\bea
{\cal B}[\delta] (z; \e ,k ,p_I)
\to
{\cal B}[\delta] (z; \e, k,p_I)
+
\sum_{i=1}^Nd_i{\cal R}_i [\delta] (z; \e , k,p_I)
\eea 
Specifically, $\R_i [\delta]$ is a form of weight $(0,0)$ in $z_i$,
and a form of weight $(1,0)\oplus (0,1)$, which is de~Rham closed, 
in each $z_j$ for $j\not=i$;
Finally, ${\cal R}_i [\delta] $ has the same monodromy as ${\cal B} [\delta] $. 

\medskip

\noindent{\bf \large (c) Holomorphicity}

\smallskip

There exist forms ${\cal S}_i [\delta]$ with properties similar
to the ones listed for ${\cal R}_i [\delta] $ so that
\bea
{\cal B}[\delta](z;\epsilon,k,p_I)
-
\sum_{i=1}^N\bar\p_i{\cal S}_i[\delta](z;\epsilon,k,p_I)
\quad \in \quad 
\otimes_{j=1}^N T_{1,0}^{z_j}(\Sigma).
\eea 
Thus the forms
${\cal B} [\delta] $ can be rewritten as
\bea
{\cal B}[\delta ] (z; \e , k,p_I)
=
{\cal H}[\delta](z; \e , k,p_I)
+
\sum_{i=1}^Nd_i{\cal S}_i [\delta] (z; \e , k,p_I)
\eea
with ${\cal H}[\delta] (z; \e , k,p_I)$ having the same monodromy as 
${\cal B}[\delta]$, of weight $(1,0)$ and holomorphic in each variable 
$z_i$ with respect to the complex structure defined by $\hat\Omega_{IJ}$, 
and away from coincident points $z_i = z_j$ for $i \not= j$.

\medskip

\noindent{\bf \large (d) Structure of the Scattering Amplitudes}

\medskip

The properties (a) and (b) imply immediately the invariance of the integrals
\bea
\int dp_I \int_{\Sigma^N}
{\cal B}[\delta] (z; \e , k,p_I)
\wedge
\overline{{\cal B}' [\bar\delta] (z; \e , k,p_I)}
\eea
under $\chi$ and $\hat\mu$ changes, thus establishing the desired gauge-slice invariance. Here, the integration is over $N$ copies 
of the worldsheet parametrized by coordinates $z_i$, $i=1,\cdots ,N$.

\medskip

The property (c) shows that the superstring amplitudes can be 
expressed as integrals over internal momenta $p_I$ and over the moduli 
space ${\cal M}_2$ of genus 2 Riemann surfaces of a Hermitian pairing of 
holomorphic forms,
\bea
\sum_{\delta,\bar\delta}{\cal C}_{\delta,\bar\delta}
\int dp_I \int_{{\cal M}_2} \int_{\Sigma ^N}
{\cal H}[\delta](z; \e , k,p_I)
\wedge
\overline{{\cal H}' [\bar\delta] (z; \e , k,p_I)}
\eea
thus restoring in this manner the basic principle of independence of left 
and right movers, as is essential to the formulations of both
the Type II and Heterotic strings. 
The coefficients $C_{\delta , \bar \delta}$ 
reflect the GSO projection, and their choices are governed by
invariance under the modular group $Sp(4, {\bf Z}_2)$.

\subsection{Summary of the Algorithm for the $N$-point Function}

\medskip

In conclusion, we have now a straightforward procedure for evaluating 
the $N$-point function, consisting of the following three steps:

\begin{itemize}

\item Deform the complex structure from the period matrix to the 
super period matrix. This requires a choice of gravitino slice
$\chiz$ and Beltrami differential $\hat \mu _{\bar z} {}^z$;

\item Insert the full vertex operator $\V$ for the emission
of each NS state. This vertex operator includes the gauge-dependent 
corrections $\V^{(1)}$ and $\V^{(2)}$.
This results in the chiral amplitude ${\cal B}[\delta]$ of (\ref{calB}) -- (\ref{Ys});

\item Extract the Dolbeault $\bar\p$-exact forms and complete 
them into de~Rham $d$-exact forms. These de~Rham exact terms 
cancel out of the superstring amplitude, upon pairing
left and right movers. The remaining forms are pure $(1,0)$ and 
holomorphic forms in each insertion point $z_i$, away from 
$z_i = z_j$ for $i \not= j$.

\end{itemize}

This procedure can be implemented for each spin structure $\delta$, 
and the gauge-fixed $N$-point function is obtained by pairing the 
holomorphic forms of the left movers with the anti-holomorphic 
forms of the right movers, and summing independently over the
spin structures $\delta,\tilde\delta$. 
The gauge-fixed $N$-point function is then guaranteed to be independent 
of all choices of gauge slice\footnote{The 
sums over spin structures can also be carried out first,
which can simplify the task of extracting the Dolbeault 
$\bar\p$-exact forms (see \cite{VI} for the case of
$N\leq 4$). The procedure for extracting Dolbeault 
$\bar\p$-exact forms for fixed $\delta$ is much more 
complicated, but it produces more holomorphic sections.}.
 
\medskip

In the present paper, we shall establish the properties (a) and (b) of \S 1.3, 
which are the ones needed for gauge-slice independence of the 
$N$-point amplitude. The explicit evaluation of the $N$-point
function and the derivation of the holomorphic blocks ${\cal H}[ \delta ]$
will be given in forthcoming publications \cite{VI, VII}.

\bigskip

Finally, we should add that there is an extensive literature on the 
calculation of $N$-point functions for the superstring \cite{vv1,old1,old4}
in the RNS formulation and more specifically on the calculation 
of two-loop amplitudes \cite{vv1,old2,old3,zwz}.
We shall discuss some of these works at greater length in the forthcoming paper 
\cite{VI} dealing with  non-renormalization theorems and explicit
formulas for the $4$-point function. We note that a different,
space-time supersymmetric, approach to superstring amplitudes 
was pursued in \cite{old0} and more recently in \cite{berkovits}.

\newpage

\section{Chiral Splitting and Chiral Blocks}
\setcounter{equation}{0}

Our considerations start from two earlier results: first, 
the superstring measure  derived in
\cite{I,II}; second, the chiral splitting theorem obtained in
\cite{dp89} for the correlation functions of scalar superfields.
We review briefly these results. For background on two-dimensional 
supergeometry, we refer to \cite{II,dp88,superg}.

\subsection{Chiral Splitting}

Let $\Sigma$ be the worldsheet, which is a surface of genus $h$ at 
perturbative order $h$, equipped with a spin structure $\delta$ and a 
supergeometry $(E_M{}^A,\Omega_M)$ satisfying the Wess-Zumino 
torsion constraints \cite{dp88,superg}. In the RNS formulation, 
superstring propagation is described  by $10$ scalar superfields 
$X^{\mu}({\bf z},\bar{\bf z})=x^\mu(z,\bar z)+\theta\psi_+^{\mu}(z,\bar z)+\bar\theta\psi_-^{\mu}(z,\bar z) + i \theta \bar \theta F^\mu (z, \bar z)$. 
The field $F^\mu$ is auxiliary and its effects are compensated by contact 
terms, as was shown in \cite{dp89}. The generating vertex for 
massless NS bosons is \cite{dp89,dpvertex},
\bea
V({\bf z},\bar{\bf z} ; \epsilon,\bar\epsilon, k)
=
{\rm exp}\,(ik^{\mu}X^{\mu}+\epsilon^{\mu}{\cal D}_+X^\mu+
\bar\epsilon^\mu{\cal D}_-\bar{X^\mu}) ({\bf z}, \bar {\bf z})
\eea
where $k^2=k\cdot\epsilon=k\cdot\bar\epsilon=0$ and ${\cal D}_\pm$ 
are  covariant derivatives with respect to the supergeometry $(E_M{}^A,\Omega_M)$. The polarization vectors $\epsilon _i$ are 
viewed as anti-commuting variables, so that $V$ is Grassmann even,
a formal device first introduced in \cite{dp89, dp88}.
The physical vertex operators  are recovered from the generating vertex 
by retaining in $V({\bf z},\bar{\bf z};\epsilon,\bar\epsilon, k) $ only the 
contribution which is linear in $\epsilon_i$, linear in $\bar \epsilon_i$ and 
integrating that part against the vertex measure 
$d^{2|2} {\bf z}_i  E({\bf z}_i)$.
The chiral splitting theorem (see specifically (4.28) and (5.4) in \cite{dp89}; 
and also \cite{dp89Rome,adp90}) 
asserts that\footnote{Throughout, we shall use the following abbreviations for 
supercurrent and stress tensor insertions,  
$\int \chi S \equiv \int \! d^2z \chiz S(z)$ and 
$\int \hat \mu T \equiv \int \! d^2 z \hat \mu_{\bar z} {}^z T(z)$.}
\bea
\label{89formula}
\<\prod_{i=1}^NV({\bf z}_i,\bar{\bf z}_i;\epsilon_i,\bar\epsilon_i, k_i)\>_X
=
\int dp_I \ \bigg | \bigg \< Q(p_I) ~  {\rm exp}\bigg ( { 1 \over 2 \pi} \int \chi S \bigg )
\prod_{i=1}^N W({\bf z}_i;\epsilon_i, k_i) \bigg \> _+  \bigg | ^2
\eea
Here, all contractions in the correlator on the right hand side are 
to be carried out with the following effective 
rules\footnote{This instruction is indicated explicitly by the $+$ subscript 
to the correlator $\< \cdots \> _+$ , but will be omitted whenever no 
confusion is expected to occur.} 
for the chiral fields $x_+(z)$ and $\psi_+(z)$,
\bea
\label{effectiverules}
\<x_+(z)x_+(w)\>_+ & = & -{\rm ln}\,E(z,w),
\no \\
\<\psi_+(z)\psi_+(w)\>_+ & = & - S_\delta(z,w),
\eea
where $E(z,w)$ and $S_\delta(z,w)$ are respectively the prime form and the Szeg\"o kernels with respect to the complex structure of $g_{mn}$ (equivalently, of $\Omega_{IJ}$). Also, 
\bea
\label{defQ}
Q(p_I) = {\rm exp} \left \{ip_I^\mu\oint_{B_I}dz\,\p_zx_+^\mu(z) \right \}
\eea
and $W$ is a chiral worldsheet superspace generating vertex, 
\bea
W({\bf z}; \e, k)
=
{\rm exp}\,\left \{ik^\mu(x_+^\mu+\theta\psi_+^\mu)(z)
+\epsilon^\mu(\psi_+^\mu+\theta\p_zx_+^{\mu})(z) \right \}
\eea
For recent applications of the subtleties of chiral splitting, see \cite{vh}.

\medskip

Next, we  need to split chirally the integrals over all supergeometries
\bea
\label{symmetric}
{\bf A} [\delta ] =\int DE_M{}^AD\Omega_M\delta(T)
\int\prod_{i=1}^N\, d^{2|2}{\bf z}_i E ({\bf z}_i)\
\<\prod_{i=1}^NV({\bf z}_i,\bar{\bf z}_i;  \e_i, \bar \e_i, k_i )\>_X,
\eea
after suitable gauge-fixing. We follow the same gauge-fixing procedure 
as in \cite{II}, \S 3. Henceforth, we assume that the genus of the worldsheet 
$\Sigma$ is $h=2$, and we fix a canonical homology basis $A_I,B_I$,
$\#(A_I\cap B_J)=\delta_{IJ}$. Supermoduli space is then of dimension 
$(3|2)$, to be parametrized by the even coordinates $\hat\Omega_{IJ}$, 
$1\leq I\leq J\leq 2$, and by two odd coordinates $\zeta^{\alpha}$,
$1\leq\alpha\leq 2$ \cite{supermoduli}. 
As in \cite{II},  \S 3, we choose metrics
$\hat g_{mn}$ whose period matrices are $\hat\Omega_{IJ}$. 
For any two independent gravitino gauge slice functions $\hat\chi_{\alpha}$, 
set
$\chi=\sum_{\alpha=1}^2\zeta^{\alpha}\hat\chi_\alpha$, and define
$\Omega_{IJ}$ by the equation (\ref{OmegaDef}), with $\chi_\alpha$ 
replaced by $\hat\chi_\alpha$. Next, choose metrics
$g_{mn}$ whose period matrices are $\Omega_{IJ}$. We may assume that $g_{mn}$ and $\hat g_{mn}$ differ only by terms of second order in 
$\zeta^{\alpha}$, so that $\hat\chi_{\alpha}$ can also be viewed as 
gravitino fields $\chi_{\alpha}$ with respect to $g_{mn}$. 
Then the supergeometry $(g_{mn},\chi=\sum_{\alpha=1}^2\zeta^{\alpha}\chi_{\alpha})$
admits $\hat\Omega_{IJ}$ as its super period matrix, and defines
a $(3|2)$-dimensional slice ${\cal S}$ for supermoduli space.
The gauge-fixing method of \cite{II}, \S 3 gives now
\bea
{\bf A} [\delta]
=
\int dp_I \int _{{\cal M}_2}
\int \prod_{i=1}^N d^{2|2}{\bf z}_i E({\bf z}_i)
\bigg|\prod_{I\leq J}d\hat\Omega_{IJ}\prod_{\alpha=1}^2d\zeta^\alpha\ 
{\cal F}[\delta](z_i, \theta _i;  \e_i, k_i, p_I)\bigg|^2
\eea
with
\bea
\label{block1}
{\cal F}[\delta]
=
{\<\prod_{a=1}^3 b(p_a)\prod_{\alpha=1}^2\delta(\beta(q_{\alpha}))\>
\over
{\rm det}\,\Phi_{IJ+}(p_a)
\
{\rm det}\,\<H_{\alpha}|\Phi_\beta^*\>}
\bigg \< Q(p_I)\, 
 {\rm exp}\bigg ( { 1 \over 2 \pi} \int \! \chi S \bigg )
\prod_{i=1}^NW({\bf z}_i, \e_i, k_i ) \bigg \>_+ \! (g)
\eea 
Here $p_a$ and $q_{\alpha}$ are arbitrary points on the surface 
$\Sigma$, $b(z)$, $\beta(z)$ are ghost and superghost fields,
and  the total supercurrent $S(z)$ is given by the sum of the matter
and ghost supercurrents, $S(z) = S_m (z) + S_{gh}(z)$, where
\bea
S_m (z) & = & - \half \p _z x_+ ^\mu \psi ^\mu _+ (z)
\no \\
S_{gh}(z) & = & \half b \gamma - {3 \over 2} \beta \p_z c - (\p_z \beta ) c
\eea
The expressions $\Phi_{IJ}=\Phi_{IJ0}+\theta\Phi_{IJ+}$, and 
$\Phi_{\beta}^*$ span a basis of superholomorphic $3/2$ differentials, while the $H_\alpha$'s are super Beltrami differentials as all prescribed in 
\cite{II}, \S 3.3-\S 3.6. The dependence on $g$ is exhibited here to stress 
that the correlation functions are taken with respect to the complex structure of $g_{mn}$.

\subsection{Deformation to $\hat\Omega_{IJ}$: the Chiral Measure}

It is important to keep in mind that in (\ref{block1}), both the finite-dimensional determinants and the correlation functions are expressed in terms of the metric 
$g_{mn}$, which depends in turn on the coordinates $(\hat\Omega_{IJ},\zeta^{\alpha})$ as well as on the choice of Beltrami 
differential $\hat\mu$ going from $\hat g_{mn}$ to $g_{mn}$,
\bea
\label{muhat}
\hat \mu_{\bar w}{}^w={1\over 2}\hat g_{w\bar w}g^{ww}
\qquad \quad
\Omega_{IJ}-\hat\Omega_{IJ}
=i\,\int\,d^2w\,\hat\mu _{\bar w} {}^w  \,\omega_I(w)\omega_J(w)
\eea
Here $w$ are holomorphic coordinates for $\hat g_{mn}$, so that 
$\hat g^{ww}=0$. The metric $g_{mn}$, and hence the Beltrami 
differential $\hat\mu$ is not unique. Different choices differ by a 
reparametrization of $\Sigma$, that is, infinitesimally by  
$\delta\hat\mu _{\bar w} {}^w =\p_{\bar w} v^w$, where 
$ v^w$ is a vector field.

\medskip

In order to carry out the integration in the odd supermoduli $\zeta^{\alpha}$ 
with $\hat\Omega_{IJ}$ as the remaining independent variables, it is 
essential to rewrite (\ref{block1})  in terms of $\hat\Omega_{IJ}$.
For the finite-dimensional determinants $\det \Phi _{IJ+} (p_a)$ and 
$\det \< H_\alpha | \Phi ^* _\beta\>$, this is done in \cite{II}, \S 5.2 and \S 6. 
In correlation functions, the shift from background metric $g_{mn}$ to $\hat g_{mn}$ can be achieved by inserting in the correlators the term 
$\int \hat\mu T$, where the total stress tensor $T(z)$ is 
the sum of the matter and ghost stress tensors, $T(z) = T_m (z) + T_{gh}(z)$,
with
\bea
T_m(z) &=& - \half \p_z x_+ ^\mu \, \p_z x_+ ^\mu  (z)
+ \half \psi ^\mu _+ \p _z \psi ^\mu _+ (z)
\no \\
T_{gh} (z) & = & 
- (\p_z b) c - 2 b \, \p_z c - {3 \over 2}  \beta \, \p_z \gamma
- {1\over 2} (\p_z \beta) \gamma
\eea
For the $0$-point function, we obtain in this way
the chiral measure  derived in (1.9) of \cite{II},
\bea
\label{measure1}  
\A  [\delta]  \!
&=&  \!
 {\< \prod _{a=1}^3 b(p_a) \prod _{\alpha=1}^2 \delta (\beta (q_\alpha)) \>
\over \det \Phi _{IJ+} (p_a) \det \< H_\alpha | \Phi ^* _\beta\>}
\biggl \{ 1 - {1 \over 8 \pi^2} \int \! d^2 \! z \chiz \! \! 
\int  \! d^2 \! w \chiw \< S(z) S(w) \>
\nonumber \\ &&  \hskip 2.6in
+ {1 \over 2 \pi} \int \! d^2 \! z  \hat \mu _{\bar z} {}^z  \< T(z)\> \biggr \}
\eea
where all correlators are now taken with respect to the metric $\hat g_{mn}$.

\subsection{Deformation to $\hat\Omega_{IJ}$: the Vertex Operators}

For the $N$-point function, upon deformation to the metric $g_{mn}$ to the metric $\hat g_{mn}$ the chiral blocks (\ref{block1}) become, 
\bea
\F  [\delta ]
=
{\< \prod _a b(p_a) \prod _\alpha \delta (\beta (q_\alpha)) \>
\over \det \Phi _{IJ+} (p_a) \det \< H_\alpha | \Phi ^* _\beta\>}
\bigg \<  \Q (p_I^\mu) \exp \bigg \{ {1 \over 2 \pi}
 \int  \big ( \chi S  + \hat \mu   T \big ) \biggr \} \prod _{i=1} ^N 
 W({\bf z}_i;\epsilon_i ,k_i) \bigg \>_+
\eea
However, in the deformation to $\hat g_{mn}$, the volume form
$d^{2|2}{\bf z}_i\ E({\bf z}_i)$ has to be deformed also. Without this 
simultaneous deformation, the correlation functions would fail to be 
gauge-invariant. In particular, the amplitudes would end
up depending on the choice of both the gravitino slice
$\chi$ and the Beltrami differential $\hat\mu$.
The deformation of the volume form $d^{2|2}{\bf z}_i\ E({\bf z}_i)$ has to be carried out with some care,
as it controls the complex type of the chiral vertex operators
-- viewed as differential forms on the worldsheet $\Sigma$ --
and ultimately their transformation properties under changes of gauge slices. 

\medskip

In components, the volume form $E({\bf z})$ for a supergeometry $(g_{mn}=e_m{}^ae_n{}^b\delta_{ab},\chi)$ on the
worldsheet $\Sigma$ is given by
(see \cite{dp88}, eqs. (3.32)-(3.33))
\bea
E = \sdet E_M {}^A
= 
{\rm det}\,(e_m{}^a)
\left (1+\half \theta\gamma^n \chi_n
+{1\over 8}\theta \bar\theta \epsilon^{mn} \chi_m \gamma_5 \chi_n \right )
\eea
where $z$ is an isothermal coordinate with respect to the metric
$g_{mn}$ and we have set auxiliary field to $0$. Under a deformation 
to the new complex structure $\hat g_{mn}$, we have 
\bea
d\xi^me_m{}^z & = & dz - \hat\mu _{\bar z} {}^z d\bar z
\no \\
{\rm det}\,(e_m{}^a) & = & 1- \hat\mu _z {}^{\bar z} \hat\mu _{\bar z} {}^z ,
\no \\
\theta\gamma^n\chi_n & = &
-\theta \hat\mu _z {}^{\bar z} \chiz - \bar\theta \hat\mu _{\bar z }{}^z \chi_z{}^-
\eea
where $z$ is now an isothermal coordinate with respect to $\hat g_{mn}$. 
Thus we obtain the key formula for the correct volume
form for the vertex operators
\bea
d^{2|2} \z ~  E ({\bf z})
=
(d \bar \theta \wedge e^{\bar z}) \, \wedge\, (d \theta \wedge e^z)
\eea
where we have set
\bea
\label{zweibein}
e^z 
= 
dz - \left (\hat \mu _{\bar z} {}^z   + \half \theta \chiz  \right )  d\bar z,
\hskip .7in
e^{\bar z} 
= 
d{\bar z} - \left (\hat \mu_z {}^{\bar z} + \half \bar \theta \chi _z ^-  \right )  d z
\eea
We stress that this decomposition is chiral, in the sense that 
each factor $(d \theta \wedge e^{z})$ or $(d \bar \theta \wedge e^{\bar z})$ depends only on $\chiz$ or $\chi_z{}^-$, but not on both. 
On the other hand, the decomposition is not holomorphic, 
in the sense that neither differential form $e^z$ or $e^{\bar z}$ 
is of pure $(1,0)$ or $(0,1)$ type. 

\medskip

This shows that the proper chiral vertex operator
should incorporate the chiral volume form and be defined by
\bea
\V  (z; \epsilon , k) 
&= & 
\int d\theta  ~ e^z ~ W ({\bf z}; \epsilon, k )
 \\
&=&
\int d\theta ~ e^z ~ \exp \bigg \{ ik ^\mu (x^\mu _+  + \theta  \psi _+) (z)
+ \e ^\mu  (\psi ^\mu _+ + \theta  \pz x^\mu _+ )(z) \bigg \}
\no \\
& = &
\e  ^\mu \bigg \{  (\pz x_+ ^\mu - i k^\nu  \psi _+ ^\mu \psi ^\nu _+ )
(dz  - \hat \mu_{\bar z} {}^z d \bar z) - \half d \bar z \chiz \psi _+ ^\mu \bigg \} 
~ e^{i k \cdot x_+ (z)}
\no 
\eea
There is an overall sign issue since the ordering of the differentials and the polarization vector $\e  ^\mu$ is a matter of convention. 
Throughout, the sign will be chosen as above. 
We note that the vertex operator $\V(z;\epsilon, k)$ is now a 1-form.
As noted earlier for $e^z$, it is however a form with both
$(1,0)$ and $(0,1)$ components.

\medskip

It will be useful to view $\V$ as a sum of terms $\V ^{(n)}$ which are of definite 
degree $n$ in the odd supermoduli $\zeta ^\alpha$, as was done in (\ref{vertexsum}), so that $\V = \V ^{(0)} + \V^{(1)} + \V^{(2)}$.
The components were given as in (\ref{vertexdecomp}), and are recalled 
here for convenience,
\bea
\label{ordervertex}
\V ^{(0)}(z; \epsilon, k)  &=& \e  ^\mu dz  (\pz x_+ ^\mu - i k^\nu  \psi _+ ^\mu \psi ^\nu _+ )(z) 
~ e^{i k \cdot x_+ (z)}
\no \\
\V^{(1)}(z; \epsilon, k) & = & - \half \e  ^\mu   d \bar z \chiz \psi _+ ^\mu (z)
~ e^{i k \cdot x_+ (z)}
\no \\
\V^{(2)} (z; \epsilon, k) & = & 
- \hat \mu _{\bar z} {}^z  ~ { d \bar z \over dz} ~ \V^{(0)} (z)
\eea
The vertex $\V^{(0)}(z; \epsilon, k)$ is of type $(1,0)$, while the vertices $\V^{(1)}(z; \epsilon, k)$ and $\V^{(2)}(z; \epsilon, k)$ are of type $(0,1)$.
Finally, we stress that all quantities in the above expression have now been 
expressed in terms of the complex structure of the superperiod matrix 
$\hat \Omega _{IJ}$. The original period matrix $\Omega_{IJ}$ no longer 
appears explicitly in our considerations; henceforth, we simply denote  $\hat\Omega_{IJ}$ by $\Omega_{IJ}$.

\medskip
Altogether, we have obtained the following formula for
the chirally symmetric contribution ${\bf A}[\delta]$ to
the superstring measure for spin structure $\delta$
\bea
{\bf A}[\delta]
=
\int_{{\cal M}_2}\int_{\Sigma^N}
\int dp_I^\mu
\ \bigg|{\cal B}[\delta](z_i;\epsilon_i,k_i,p_I^\mu)\bigg|^2,
\eea
where ${\cal B}[\delta]$ are the $1$-forms in each $z_i$
defined in (\ref{calb}).

\subsection{Non-Renormalization of the Super-Period Matrix}

The simple contraction, using (\ref{effectiverules}), of the operator $Q(p_I)$, 
ignoring vertex operators and the insertions of the stress tensor and the supercurrent, gives rise to the following factor which is familiar from chiral splitting,
\bea
\left \< Q(p_I)\, \right \>  = \exp \{ i \pi p^\mu _I \hat \Omega _{IJ} p^\mu _J \}
\eea
Further contractions of $Q(p_I)$ with the stress tensor and supercurrent 
insertions  produce corrections to this Gaussian.
Actually, these corrections cancel one another, as is indeed expected to happen for consistency.  The respective corrections are given as follows. 
For the insertion of the stress tensor, the contribution is given by,
\bea
\label{Tcont}
\left \<  {1 \over 2 \pi} \int \hat \mu T  
~ Q(p_I) \right \> _{(c)}
& = & 
\half \left \< Q(p_I) \right \> 
p^\mu _I p^\mu _J \oint _{B_I} \! \! dz  \oint _{B_J} \! \! dw \int \! {d^2 u \over 2 \pi}
\hat \mu _{\bar u} {}^u  \pz \pu \ln E ~ \pz \pw \ln E
\no \\ 
& = & - \pi p^\mu _I p^\mu _J \left \< Q(p_I) \right \>
\int \hat \mu  \omega _I  \omega _J
\no \\ 
& = & -i \pi p^\mu _I p^\mu _J \left \< Q(p_I) \right \>
\left ( \hat \Omega _{IJ} - \Omega _{IJ} \right )
\eea
where we have used the effective rule of (\ref{effectiverules}) for the contraction
of chiral bosons. Also, in passing from the second to the last
line, we have used the definition of $\hat \mu$, in (\ref{muhat}).
For the insertion of two supercurrents, the contribution is given by
\bea
&&
\left \< \half \bigg ( {1 \over 2 \pi} \int \! \chi S  \bigg )^2 Q(p_I) \right \>_{(c)}
\eea
Here, no contractions of both $x_+$ and $\psi_+$
are allowed between the two supercurrents, since such a term already
belongs to the disconnected part.
Performing first the $\psi_+$ contraction, using (\ref{effectiverules}),
we find the following expression,
\bea
\left \<  \bigg ( {1 \over 2 \pi} \int \chi S  \bigg )^2 \right \> _{\psi _+}
=
- {1 \over 4 \pi ^2} \int d^2 u  \int d^2 v 
\p x_+ ^\sigma (u) \chiu S_\delta (u,v) \chiv \p x_+ ^\sigma (v)
\eea
Performing now also the contractions of $x_+$, using (\ref{effectiverules}),
we find
\bea
\label{SScont}
&&
\left \< \half \bigg ( {1 \over 2 \pi} \int \chi S  \bigg )^2 Q(p_I) \right \> _{(c)}
\no \\  
&& 
\qquad
= 
- {p^\mu _I p^\mu _J \over 32 \pi ^2} \< Q(p_I)\>
 \oint _{B_I} \! \! dz  \oint _{B_J} \! \! dw 
\int \! d^2 u \int \! d^2 v \chiu  S_\delta (u,v) \chiv
\p_z \p_u \ln E ~ \p _w \p _v \ln E
\no \\  
&& \qquad
= 
{1 \over 8} p^\mu _I p^\mu _J \< Q(p_I)\> \int \! \! d^2u  \int \! \! d^2v 
~ \omega _I (u) \chiu S_\delta (u,v) \chiv \omega _J (v) 
\no \\  
&& \qquad
= 
\pi i p^\mu _I p^\mu _J \< Q(p_I)\> \left ( \hat \Omega _{IJ} - \Omega _{IJ} \right )
\eea
In the passage to the last line, we have used the definition of $\hat \Omega_{IJ}$
in (\ref{OmegaDef}). Clearly, the contributions from the insertion of $T$ in 
(\ref{Tcont}) and of $SS$ in (\ref{SScont}) cancel one another.

\newpage

\section{The $dx_+^\mu$ Formulation of Vertex Operators}
\setcounter{equation}{0}

The conceptual difficulties associated with the fact that the
full vertex operator $\V$ is now gauge-slice dependent have been
stressed in the Introduction, section \S 1.
The gauge-slice dependent corrections $\V^{(1)}$ and $\V^{(2)}$
give rise to some practical difficulties as well. In practice,
it is often convenient to choose Dirac point masses for the gravitino 
slice $\chiz$ and the Beltrami differential $\hat\mu _{\bar z} {}^z$. 
How to do so when $\chiz$ and $\hat\mu _{\bar z} {}^z$ appear in 
an ``unintegrated" form in a chiral amplitude, as in $\V^{(1)}$ and $\V^{(2)}$,
is fraught with difficulties, and can easily lead to contradicting outcomes.

\medskip
In this section, we show that there is an alternate formulation
of the full vertex operators $\V$, where the unintegrated $\chiz$ cancel out. 
This shows that, when evaluating the $N$-point function, it is safe to 
take $\chiz$ to consist of Dirac point masses at the outset. 
However, the Beltrami differential $\hat\mu _{\bar z} {}^z$ still 
appears in an unintegrated form,
and it is not reliable to take it as consisting of Dirac point masses. These basic guidelines will be followed in the explicit evaluation of the $N$-point function for $N\leq 4$ in \cite{VI}.

\medskip

The key to the alternative formulation   is
a natural geometric object $\U (z)$, defined by
\bea
\U  (z; \epsilon, k) 
\equiv \e ^\mu \left ( dx_+ ^\mu - i k^\nu dz ~ \psi _+ ^\mu \psi _+ ^\nu
\right )(z)  ~ e^{ik \cdot x_+ (z) }
\eea
Here, $d$ is the differential operator acting on the point of the vertex $z$.
This vertex differs from $\V ^{(0)}$ as follows,
\bea
\U (z; \epsilon, k) 
& = & 
\V ^{(0)} (z; \epsilon, k) + \K  (z; \epsilon, k)
\no \\
\K (z; \epsilon, k) 
& = & 
\e ^\mu ~ d\bar z ~ \p _{\bar z} x_+ ^\mu (z)~ e^{i k \cdot x_+(z)}
\eea 
We now wish to re-express the amplitude in terms of the  vertices $\U$.
The difference between the vertices $\V^{(0)}(z_i; \epsilon_i, k_i)$ and 
$\U (z_i; \epsilon_i, k_i)$ involves a  $\p _{\bar z} x_+(z_i)$ factor which, 
upon contraction of the field $x_+$, produces contributions at the points 
where $x_+$ is not complex analytic. 
There are four instances when non-analyticities appear,
\begin{enumerate}
\item at insertion points $z_j$, different from $z_i$; 
\item at the insertion points of the supercurrent $S$;
\item at the insertion of the stress tensor $T$;
\item at the insertion of the internal momentum operator in $Q(p_I)$.
\end{enumerate}
The effects of 1 are immaterial in view of the ``cancelled propagator 
argument", i.e. the analyticity as a function of external momenta. 

\subsection{Preliminaries}

Thus, we need to obtain only the contributions from 2, 3, and 4. 
These are computed using the following contraction rules, where
only the factor $\p _{\bar z} x_+ ^\mu (z)$ is contracted but all 
other fields are left for contraction at a later stage. We denote
this procedure by $\< \cdots \>_*$. We have,
\bea
\label{prel}
\left \< { 1 \over 2 \pi} \int \!  \chi S ~ 
\p _{\bar z} x_+ ^\mu (z) \right \>_*
& = &
- \half \chiz \psi _+ ^\mu (z)
\\
\left \< { 1 \over 2 \pi} \int \!  \hat \mu  T ~ 
\p _{\bar z} x_+ ^\mu (z) \right \>_*
& = &
-  \hat \mu _{\bar z} {}^z \p  _z x_+ ^\mu (z)
\no \\
\left \< Q(p_I)\, \p _{\bar z} x_+ ^\mu (z) \right \>_* 
& = & 
Q(p_I)\, i p_I ^\mu \p _{\bar z} \oint _{B_I} \! dw \< \p _w x_+ (w)  x_+ (z) \> =0
\no
\eea
From the last line, it follows that contractions of item 4 above do not occur.
It is useful to have the contraction of the entire operator $\K$,
using the same rules and the same notations. After a brief calculation,
using (\ref{prel}), we find $\< Q(p_I)\, \K  _i  \>_* = 0$ and
\bea
\left \< { 1 \over 2 \pi} \int \! \chi S ~ \K _i \right \>_*
& = &
\V ^{(1)} _i
\no \\
\left \< { 1 \over 2 \pi} \int \! \hat \mu  T ~ \K _i  \right \>_*
& = &
\V ^{(2)} _i - \L _i
\eea
where $\L_i$ is given by
\bea
\L_i = 
i \hat \mu _{\bar z _i} {} ^{z_i}  
d\bar z_i  ~ \e _i ^\mu k_i ^\nu  \psi _+ ^\mu \psi _+ ^\nu (z_i)
~ e^{i k _i \cdot x_+ (z_i)}
\eea
This contribution has a natural interpretation. The 
product $\psi _+ ^\mu \psi _+ ^\nu$ is the $z$-component
of a local Lorentz vector, so that the Lorentz invariant
$(dz - \hat \mu d\bar z) \psi _+ ^\mu \psi _+ ^\nu$
is actually the proper 1-form.

\subsection{Reformulation}

We reformulate $\Y_1, \Y_2, \Y_3, \Y_4, \Y_5$ in terms of the 
vertices $\U_i$. Substitution yields,
\bea
\Y _1 & = &
{1 \over 8 \pi ^2} \bigg \< Q(p_I)\, \int \! \chi S  \int \! \chi S ~ 
\prod _{i=1}^N \U _i  \bigg \> _{(c)}
- \sum _{i=1} ^N {1 \over 8 \pi ^2} 
\left \< Q(p_I)\, \int \! \chi S  \int \! \chi S  ~ \K _i \prod _{j \not= i}^{N} \U _j
\right \> _{(c)}
\no \\ &&
+  \half \sum _{i \not= j} ^N {1 \over 8 \pi ^2} 
\left \< Q(p_I)\, \int \! \chi S  \int \! \chi S    ~ \K  _i  ~ \K _j  
\prod _{l \not= i,j} ^N \U_l \right \> _{(c)}
\eea
Using the $\< \cdots \>_*$ contractions, we obtain,
\bea
\Y _1 & = &
{1 \over 8 \pi ^2} \left \< Q(p_I)\, \int \! \chi S ~ \int \! \chi S ~ 
\prod _{i=1}^N \U _i \right \> _{(c)}
- \sum _{i=1} ^N {1 \over 2 \pi } 
\left \< Q(p_I) \, \int \! \chi S     ~ \V _i ^{(1)} \prod _{j  \not= i}^{N}  \U _j \right \>
\no \\ &&
+  \half \sum _{i \not= j} ^N  \left \< Q(p_I)   ~ \V _i ^{(1)}   ~ \V _j  ^{(1)}
\prod _{l \not= i,j}^N  \U _l \right \>
\eea
The remaining contributions are calculated in a similar manner
and are given by
\bea
\Y _2 
& = & 
{1 \over 2 \pi} \left \< Q(p_I)\, \int \! \hat \mu T ~ 
\prod _{i=1}^N \U _i \right \> _{(c)}
+ 
\sum _{i=1} ^N  \left \< Q(p_I)\,  
  \left ( -\V _i ^{(2)} + \L _i \right )  \prod _{j  \not= i }^N \U_j \right \>
\no \\
\Y _3 & = & 
{1 \over 2 \pi} \sum _{i=1} ^N 
\left \< Q(p_I)\, \int \! \chi S  ~ \V ^{(1)} _i  \prod _{j \not= i}^N  \U _j \right \>
-  \sum _{i \not= j j } ^N  \left \< Q(p_I) ~ \V _i ^{(1)}  ~ \V _j ^{(1)} 
\prod _{l \not= i,j}^N \U _l \right \>
\no \\
\Y _4 & = &
\half \sum _{i  \not= j } ^N  \left \< Q(p_I)  ~ \V _i ^{(1)}  ~ \V _j ^{(1)} 
\prod _{l \not= i,j}^N \U _l \right \>
\no \\
\Y _5 
& = & 
\sum _{i=1} ^N  \left \< Q(p_I)  ~ \V _i ^{(2)} \prod _{j  \not= i}^N \U _j
\right \>
\eea
Putting all together, we obtain our final result,
\bea
\label{Urepresentation}
\sum _{a=1} ^5 \Y_a 
& = &
{1 \over 8 \pi ^2} 
\left \< Q(p_I)\, \int \! \chi S  \int \! \chi S ~ \prod _{i=1}^N \U _i \right \> _{(c)}
 \\ &&
+
 {1 \over 2 \pi}  \left \< Q(p_I)\,  \int \hat \mu T ~ \prod _{i=1}^N \U _i \right \> _{(c)}
+
 \sum _{i=1} ^N  \left \< Q(p_I)  ~ \L _i    \prod _{j \not= i}^{N}\right \>
\no
\eea
Clearly, in terms of the vertex $\U$, the amplitude simplifies 
considerably.

\newpage

\section{The Amplitudes ${\cal B}[\delta]$
as Closed Differential Forms}
\setcounter{equation}{0}

The first important property of the vertex operators $\V_i$ is that their 
correlation functions are closed differential forms
with respect to each $z_i$. This property is manifest for the disconnected 
part $\B [\delta]^{(d)} $ since the correlator $\< Q(p_I) \prod _i \V ^{(0)} \>$
is a holomorphic 1-form in each $z_i$ away from coincident points.
For the connected part, the closedness is equivalent to the equation, 
\bea
d_i\sum_{a=1}^5\Y_a=0,
\ \ \ 1\leq i\leq N.
\eea
There are at least two ways of seeing this. The first way makes
use of the superholomorphicity of the chiral blocks. The second
way relies on the representation (\ref{Urepresentation}) of the $\Y_a$ in terms of the vertices $\U_i$ obtained in \S 2.6.
We discuss them both.

\medskip

In the first way, we observe that the closedness property of a differential form is a notion which depends only on the $C^\infty$ structure, and not on the complex structure. Thus we can verify it in the original conformal coordinates $z$ of the metric $g_{mn}$, and do not need to deform to $\hat\Omega_{IJ}$
with the Beltrami differential $\hat\mu_{\bar z}{}^z$. The chiral
volume form is then 
$d\theta\wedge e^z=d\theta\wedge (dz-{1\over 2}\theta\chiz d\bar z)$, and it suffices to show that the expression
\bea
\label{1form}
\int d\theta\wedge (dz-{1\over 2}\theta\chiz d\bar z)
{\cal F}[\delta](z,\theta,{\bf z}_2,\cdots,{\bf z}_N)
=
dz\,{\cal F}_+ [\delta ] -{1\over 2}d\bar z\,\chiz {\cal F}_0 [\delta]
\eea
is a closed $1$-form in $z$. Here we have set $(z_1,\theta_1)=(z,\theta)$,
${\cal F}[\delta]={\cal F}_0 [\delta ] +\theta {\cal F}_+ [\delta] $
to simplify  notation, the other variables ${\bf z}_i$ being treated 
successively in the same way. 
However, the chiral blocks ${\cal F}[\delta]$ are superholomorphic, that is,
The chiral blocks ${\cal F}[\delta]$ are superholomorphic, i.e.,
\bea
\label{superholomorphicity}
{\cal D}_-{\cal F}[\delta]=0,
\qquad {\bf z_i}\not={\bf z}_j.
\eea  
Decomposing this superholomorphicity 
property into components, we have
\bea
{\cal D}_-{\cal F}[\delta]
=
\bar\theta (\p_{\bar z}{\cal F}_+ [\delta] +{1\over 2}\chiz{\cal F}_+ [\delta] )
+
\bar \theta \theta (\p_{\bar z}{\cal F}_+ [\delta] 
+{1\over 2}\p_z(\chiz{\cal F}_0) [\delta] )
\eea 
Thus, the holomorphicity condition ${\cal D}_-{\cal F}[\delta]=0$ implies that $\p_{\bar z}{\cal F}_+ [\delta] +{1\over 2}\p_z(\chiz{\cal F}_0 [\delta])=0$, which is exactly the condition that
the $1$-form (\ref{1form}) be closed.

\medskip  
 
The second way is based on the conformal coordinates $z$
for the metric $\hat g_{mn}$. It is slightly longer, but it
illustrates well the roles of the various terms in
the representation (\ref{Urepresentation}).
The exterior derivative of the amplitudes with respect to any one of the vertex insertion points is computed from the exterior derivative of the individual vertex operators, viewed as differential forms.
The differential of the individual vertex is easily obtained,
\bea
\label{dvertex}
d \U  =
 \e ^\mu  k^\nu 
\left \{ - i dx_+ ^\mu  \wedge dx_+ ^\nu 
    + k^\sigma   d x_+ ^\sigma  \wedge dz ~ \psi ^\mu _+ \psi ^\nu _+  
    - i  d \left ( dz ~ \psi ^\mu _+ \psi ^\nu _+ \right ) \right \} 
    ~ e ^{i k \cdot x_+ }
\eea

\medskip

We begin by focussing on the $\p _{\bar z}$ parts in $d \U$, and their 
contractions inside the amplitude. By the ``cancelled propagator" argument,
any contractions of the $\p _{\bar z}$ terms with fields in other
vertex operators will yield $\delta (z_i,z_j)$ contact terms which,
by external momentum analyticity, must vanish. Thus, the $\p _{\bar z}$
parts of $d \U$ must be contracted only with the operators $S$ and $T$.
Preliminary formulas that will come in handy are as follows,
\bea
\label{prel2}
\left \< { 1 \over 2 \pi} \int \!  \hat \mu  T ~ 
\p _{\bar z} x_+ ^\mu (z) \right \>_*
& = &
-  \hat \mu _{\bar z} {}^z \p  _z x_+ ^\mu (z)
\no \\
\left \< { 1 \over 2 \pi} \int \! \hat \mu T ~ 
\p _{\bar z} (\psi ^\mu _+ \psi _+ ^\nu ) \right \>_*
& = &
-  \p _z \left (\hat \mu _{\bar z} {}^z \psi _+ ^\mu \psi _+ ^\nu \right ) (z)
\eea
which may be derived using the standard Green functions
for $x_+$ and $\psi _+$. These expressions
are intimately related with the corresponding diffeomorphism 
transformations of these fields (see e.g. \cite{dp88},
\S III.C).

\subsection{$\hat \mu$-Dependence}

The first contraction we need to evaluate is
\bea 
\label{muvertex}
\left \< {1 \over 2 \pi} \int  \hat \mu  T  ~ d\U(z) \right \> _*
\eea
Using (\ref{prel2}), the first term of (\ref{dvertex}) contracts as follows,
\bea
&&
\left \< {1 \over 2 \pi} \int \hat \mu  T  ~ 
\left ( \p_z x_+ ^\mu \p _{\bar z} x_+ ^\nu - \p_z x_+ ^\nu \p _{\bar z} x_+ ^\mu 
\right ) \right \> _* 
= 
- \hat \mu _{\bar z} {}^z \p_z x_+ ^{[\mu} \p _z x_+ ^{\nu ]} 
=0
\eea
and therefore does not contribute to (\ref{muvertex}). The second and third 
terms in (\ref{dvertex}) contract as follows,
\bea
\left \< {1 \over 2 \pi} \int \!  \hat \mu  T ~
\p _{\bar z} x_+ ^\sigma \psi ^\mu _+ \psi _+ ^\nu e^{i k \cdot x_+} \right \> _*
& = &
- \hat \mu _{\bar z} {}^z \p _z x_+ ^\sigma \psi ^\mu _+ \psi _+ ^\nu e^{i k \cdot x_+}
\no \\ 
\left \< {1 \over 2 \pi} \int \!  \hat \mu  T ~ \p _{\bar z} 
\left ( \psi ^\mu _+ \psi _+ ^\nu \right ) e^{i k \cdot x_+} \right \>_* 
& = &
- \p _z \left ( \hat \mu _{\bar z} {}^z 
\psi ^\mu _+ \psi _+ ^\nu (z) \right ) e^{i k \cdot x_+}
\eea
Combining both of these, we get
\bea
\left \< {1 \over 2 \pi} \int \!  \hat \mu  T ~
\left [ i k^\sigma \p _{\bar z} x_+ ^\sigma \psi ^\mu _+ \psi _+ ^\nu 
+ \p _{\bar z}  \left ( \psi ^\mu _+ \psi _+ ^\nu \right ) 
\right ] e^{i k \cdot x_+} \right \> _*
=
- \p_z \left ( \hat \mu _{\bar z} {}^z   \psi ^\mu _+ \psi _+ ^\nu e^{i k \cdot x_+} \right )
\eea
and further combining with the momenta and polarization vector, we obtain,
\bea
\left \< {1 \over 2 \pi} \int \!  \hat \mu  T ~ d_i \U_i (z_i) \right \> _*
=
- i \e ^\mu_i  k ^\nu_i  dz_i \wedge d\bar z_i 
\p_{z_i} \left ( \hat \mu _{\bar z _i} {}^{z_i}   \psi ^\mu _+ \psi _+ ^\nu (z_i) 
e^{i k \cdot x_+} \right )
\eea
Recalling now that there is also an additional contribution to the 
amplitude denoted by $\L_i$, which produces the following derivative term,
\bea
d_i \L_i = 
i \e _i ^\mu k_i ^\nu dz_i \wedge d \bar z _i \p _{z_i} \left ( 
\hat \mu _{\bar z_i} {}^{z_i} ~   \psi _+ ^\mu \psi _+ ^\nu (z_i)
~ e^{i k _i \cdot x_+ (z_i)} \right )
\eea
we see that the sum of all contributions involving $\hat \mu$ cancel
as follows, 
\bea
\left \< {1 \over 2 \pi} \int \!  \hat \mu  T ~ d_i \U_i (z_i) \right \> _*
+ d_i \L_i =0
\eea
Hence, all contractions of the $\p _{\bar z}$ terms with the stress tensor vanish. 

\subsection{$\chi$-Dependence}

The $\chi$-dependence governs the contractions of the 
$\p _{\bar z}$-terms with the supercurrent $S$, and we 
make use of the following preliminary formulas,
\bea
\left \< { 1 \over 2 \pi} \int \!  \chi S ~ 
\p _{\bar z} x_+ ^\mu (z) \right \>_*
& = &
- \half \chiz \psi _+ ^\mu (z)
 \\
\left \< { 1 \over 2 \pi} \int \!  \chi S ~ 
\p _{\bar z} \left ( \psi _+ ^\mu \psi _+ ^\nu \right ) \right \>_*
& = &
- \half \chiz \left ( \p _z x_+ ^\mu \psi _+ ^\nu 
-  \p _z x_+ ^\nu \psi _+ ^\mu \right ) (z)
\no
\eea
which are intimately connected with 
local worldsheet supersymmetry transformations. We need to 
evaluate the following contractions,
\bea
\left \< {1 \over 2 \pi} \int  \chi  S  ~ d\U(z) \right \> _*
\eea
The three parts of $d \U$ given rise to the following contractions,
\bea
\left \< {1 \over 2 \pi} \int \!  \chi  S  ~ 
\left ( \p_z x_+ ^\mu \p _{\bar z} x_+ ^\nu - \p_z x_+ ^\nu \p _{\bar z} x_+ ^\mu 
\right ) e^{i k \cdot x_+} \right \> _*
& = & 
- \half \chiz \left ( \p _z x_+ ^\mu \psi _+ ^\nu -\p _z x_+ ^\nu \psi _+ ^\mu \right )
e^{i k \cdot x_+}
\no \\
\left \< {1 \over 2 \pi} \int \!  \chi S ~
\p _{\bar z} x_+ ^\sigma \psi ^\mu _+ \psi _+ ^\nu e^{i k \cdot x_+} \right \> _*
& = &
- \half \chiz   \psi ^\sigma _+ \psi ^\mu _+ \psi _+ ^\nu e^{i k \cdot x_+}
\\ 
\left \< {1 \over 2 \pi} \int \!  \chi S ~
\p _{\bar z} 
\left ( \psi ^\mu _+ \psi _+ ^\nu \right ) e^{i k \cdot x_+} \right \> _*
& = &
- \half  \chiz \left ( \p _z x^\mu _+ \psi _+ ^\nu
- \p _z x_+ ^\nu \psi _+ ^\mu \right ) e^{i k \cdot x_+}
\no
\eea
Inserting the factors of momenta and polarization vectors in 
the second line gives,
\bea
\left \< {1 \over 2 \pi} \int \!  \chi S ~
\e ^\mu k^\nu k ^\sigma 
\p _{\bar z} x_+ ^\sigma \psi ^\mu _+ \psi _+ ^\nu e^{i k \cdot x_+} \right \> _*
=
- \half \e ^\mu \chiz   (k \cdot \psi  _+) \psi ^\mu _+ (k \cdot \psi _+ ) 
e^{i k \cdot x_+} =0 \quad 
\eea
The cancellation occurs because of the square of the single fermionic
field factor vanishes, $(k \cdot \psi  _+)(k \cdot \psi _+ ) =0$.
Using the cancellation generated by the second line, the first 
and third lines combine to yield 0 to the correlator of $d \U$,
so that we have,
\bea
\left \< {1 \over 2 \pi} \int \!  \chi S ~ d_i\U_i (z_i) \right \> _*
& = & 0
\eea
Therefore, all contractions of all $\p _{\bar z}$ terms in $d\U$
with the supercurrent vanish as well.
The closedness of the differential form
$\sum_{a=1}^5\Y_a$ is established.

\newpage

\section{Gauge Slice Variations as Exact Differentials}
\setcounter{equation}{0}

In this section, we prove that the full chiral amplitudes, including the chiral 
contribution of the volume form, transform in a simple manner under 
changes in slice for the Beltrami differentials $\hat \mu$ and $\chi$. 
The changes of the chiral  amplitudes are by the addition of an exact 
differential of a well-defined and single-valued differential in the external 
insertion points. 

\subsection{Dependence on the $\hat \mu$-Slice}

Dependence on the $\hat \mu$-slice is investigated using a variation 
of the slice according to the following rule,
which leave $\hat\Omega_{IJ}$ invariant
\bea
\delta _v \hat \mu_{\bar z} {}^z (z) & = & \p_{\bar z} v^z (z)
\no \\
\delta _v \chiz (z) & = & 0
\eea
Here,  $v^z$ is a single-valued, non-singular vector field. Also,
since $\hat \mu$  is bilinear in the odd supermoduli $\zeta ^\alpha$, 
$v^z$ must  be bilinear in $\zeta ^\alpha$. 

\medskip

The variation of $\hat \mu$ affects the amplitude $\B[\delta] $ as follows.
The variation of the {\sl disconnected part}, $\B [\delta] ^{(d)} $ has already 
been carried out in \cite{II} and cancels. The variation of the {\sl connected part}
may be computed from the variations of the terms in $\B [\delta ] ^{(c)}$. 
To determine the latter, we need the variations of the
vertex operators, which are given by
\bea
\delta _v \V ^{(0)}  (z) & = & 0
\no \\
\delta _v \V ^{(1)}  (z) & = & 0
\no \\
\delta _v \V ^{(2)}  (z) & = & - \p _{\bar z} v^z {d \bar z \over dz} \V ^{(0)} (z)
\eea
As a result, we have $\delta _v \Y_1= \delta _v \Y_3 = \delta _v \Y_4=0$.
Notice that we also have $\delta _v (d \mu _0 [\delta] )=0$, in view of the results 
of paper II. The variations that remain are as follows,
\bea
\delta _v \Y _2 
& = & 
{1 \over 2 \pi}
\int \! d^2 z \p_{\bar z} v^z \left \< Q(p_I) \,  T(z)  ~ \prod _{i=1} ^N \V_i^{(0)}
\right \> _{(c)}
\no \\
\delta _v \Y _5
& = & 
-  \sum _{i =1}^N \p _{\bar z_i} v^{z_i} (z_i) {d \bar z_i \over dz_i}   
\left \< Q(p_I) \, \prod _{j =1}^N \V_j ^{(0)} \right \> 
\eea
The first variation may be computed using the following  standard OPEs,
\bea
T(z) ~ \p_w x_+ (w) 
& = & 
{1 \over (z-w)^2} \p_w x_+ (w) + {1 \over z-w} \p_w ^2 x_+(w) + \O (1)
\no \\
T(z) ~ \psi ^\mu _+ \psi ^\nu _+ (w) 
& = & 
{1 \over (z-w)^2} \psi ^\mu _+ \psi ^\nu _+ (w) + 
{1 \over z-w} \p_w ( \psi ^\mu _+ \psi ^\nu _+) (w)  + \O (1)
\eea
and hence
\bea
{1 \over 2 \pi} \int \! d^2 z \p _{\bar z} v^z T(z) \, \p_w x_+(w) 
& = & - \p_w \bigg (v^w \p_w x_+(w) \bigg )
\no \\
{1 \over 2 \pi} \int \! d^2 z \p _{\bar z} v^z   T(z)\,  \psi ^\mu _+ \psi ^\nu _+ (w) 
& = & - \p_w \bigg ( v^w \psi ^\mu _+ \psi ^\nu _+ (w) \bigg  )
\eea
From these relations, we deduce the following useful identities,
\bea
{1 \over 2 \pi} \int \! d^2 z \p _{\bar z} v^z  T(z) \,  Q(p_I) \, 
& = & - p^\mu _I \oint _{B_I} dw ~ \p_w (v^w \p_w x_+ (w)) Q(p_I) \,
=  0
\no \\
{1 \over 2 \pi} \int \! d^2 z \p _{\bar z} v^z  T(z) \, \V ^{(0)} (w) 
& = & - \p_w \bigg ( v^w \V ^{(0)} (w) \bigg  )
\eea
Applying these results to the calculation of $\Y_2$, we find,
\bea
\delta _v \Y_2 & = & 
- \sum _{i=1} ^N \p_{z_i} \left (
v^{z_i} \left \< Q(p_I) \, \prod _{j=1} ^N \V^{(0)} _j \right \> \right )
\eea
The second variation may be recast as follows,
\bea
\delta _v \Y _5
& = &
-  \sum _{i =1}^N  {d \bar z_i \over dz_i}  \p _{\bar z_i} \left ( v^{z_i} (z_i)  
\left \< Q(p_I) \, \prod _{j =1}^N \V_j ^{(0)}  \right \>  \right )
\no \\ && 
+  \sum _{i =1}^N  {d \bar z_i \over dz_i} v^{z_i} (z_i)   \p _{\bar z_i} 
\left \< Q(p_I) \, \prod _{j =1}^N \V_j ^{(0)} \right \>  
\quad
\eea
As it is assumed throughout that the points $z_i$ are distinct, i.e.
$z_i \not= z_j$ whenever $i \not= j $, the second term above 
vanishes since away from the coincident locus, these correlators
are holomorphic in $z_i$. Combining both variations, and recognizing 
the components of the exact differential 
$ d_i \equiv dz_i \p _{z_i} + d\bar z _i \p _{\bar z_i}$,
we have,
\bea
\delta _v \B [\delta ]^{(c)}  = d\mu _0 [\delta] (\delta _v \Y _2 + \delta _v \Y _5)
= \sum _{i=1}^N d_i \R_{i} ^v [\delta] 
\eea
Here, the differential forms $\R_{i}^v [\delta] $  are given by
\bea
\R_i ^v  [\delta ] =
- d\mu _0 [\delta]  v^{z_i} (dz_i)^{-1} \left \< Q(p_I) \, \prod _{j =1}^N \V_j ^{(0)}  \right \> 
\eea
This form has the following properties,
\begin{itemize}
\item $ \R_i ^v[\delta]  $ is a form of type $(1,0)$ in $z_j$ for $j \not= i$;
\item $ \R_i ^v[\delta]  $ is a form of type $(0,0)$ in $z_i$;
\item $ \R_i ^v[\delta]  $ is holomorphic in $z_j$ for $j\not= i$.
\end{itemize}
This concludes the proof of our assertion on $\hat \mu$-slice dependence.

\subsection{Dependence on the $\chi$-Slice}

The $\chi$-slice variation is given by the following rule,
\bea
\delta _\xi \mu _{\bar z} {}^z (z) & = & \xi ^+  \chiz (z)
\no \\
\delta _\xi \chiz (z) & = & - 2 \p _{\bar z} \xi ^+ (z)
\eea
Since $\chi$ itself is linear in the odd supermoduli $\zeta ^\alpha$, 
$\xi ^+$ is also linear in $\zeta ^\alpha$. Furthermore, $\xi ^+$ is a 
single-valued, non-singular spinor field.

\medskip

The variation of $\chi$ affects the amplitude as follows. 
The variation of the  {\sl disconnected part}, $\B [\delta ] ^{(d)}$, 
has already been carried out in \cite{II} and cancels. 
The variation of the {\sl connected part} may be computed
from the variations of the terms in $\B [\delta ] ^{(c)} $. 
(Notice that we  have $\delta _\xi (d \mu _0[\delta]  )=0$ automatically.)
To determine these, we need the variations of the vertex operator pieces, 
which are given by,
\bea
\delta _\xi \V ^{(0)}  (z) & = & 0
\no \\
\delta _\xi \V ^{(1)}  (z) & = & - d\bar z \left  (\p_{\bar z} \xi ^+ (z) \right ) 
\e ^\mu \psi ^\mu _+ (z) e^{i k \cdot x_+(z)}
\no \\
\delta _\xi \V ^{(2)}  (z) & = & - \xi ^+  \chiz  {d\bar z \over dz} \V ^{(0)} (z)
\eea
Using these ingredients, we can recast the variations of the $\Y$'s in a more
explicit form,
\bea
\delta _\xi \Y _1 
& = & 
 {1 \over 4 \pi ^2} \left \< Q(p_I) \, 
\int \! d^2 z ~ \delta _\xi \chiz (z) S(z) \int \! \chi S
~ \prod _{i=1}^N \V_i ^{(0)}  \right \> _{(c)}
\no \\
\delta _\xi \Y _2 
& = & 
{1 \over 2 \pi } \left \< Q(p_I) \, \int \! d^2 z ~ \xi ^+ (z) \chiz (z) T(z) 
~ \prod _{i=1}^N \V_i ^{(0)}  \right \> _{(c)}
\no \\
\delta _\xi \Y _3 
& = & 
{1 \over  2 \pi } \sum _{i=1} ^N \left \< Q(p_I) \, 
\left \{ \int \! d^2 z ~ \delta _\xi \chiz (z)  S(z) ~  \V ^{(1)} _i  + 
\int \! \chi S ~  \left (\delta _\xi \V ^{(1)} _i  \right )   \right \}
\prod _{j  \not= i}^N  \V_j ^{(0)}  \right \> 
\no \\
\delta _\xi \Y _4 
& = & 
\sum _{i \not= j} \left \< Q(p_I) ~ \left (\delta _\xi \V^{(1)} _i  \right )  ~  \V^{(1)} _j  
\prod _{l  \not= i,j}^N  \V_l ^{(0)} 
\right \> 
\no \\
\delta _\xi \Y _5 
& = & 
- \sum _{i =1}^N \xi ^+ \chiz (z_i) \left \< Q(p_I) \,
 \prod _{j =1}^N \V_j ^{(0)}  \right \> 
\eea

\subsubsection{Preliminaries}

We begin with the evaluation of $\delta _\xi \Y_1$. We make use of the 
following OPEs,
\bea
S(z)  x_+ (w) & = & -\half {1 \over z-w} ~ \psi _+ (w)
\no \\
S(z) \psi _+ (w) & = & - \half {1 \over z-w} ~ \p_w x_+(w)
\eea
Of course, $x_+(w)$ in the first line is not a well-defined 
conformal field (but $\p_w x_+(w)$ is). The OPE given here
is to be understood as used in differences and under
derivatives, in which  $x_+$ effectively becomes a conformal 
field and the OPE is well-defined. 
The action of the integrated supercurrent 
on bosonic and fermionic fields is by derivation and 
therefore fixed by the following elementary rules of
operation,
\bea
\delta _\xi \left \< {1 \over 2 \pi} \int \! \chi S ~x_+(w) \right \>_x
& \equiv & \delta _{ ss} x_+(w) =  \xi ^+ \psi _+(w)  
\no \\
\delta _\xi \left \< {1 \over 2 \pi} \int \! \chi S ~\psi_+(w) \right \>_\psi
& \equiv & \delta _{ ss} \psi _+(w) =  \xi ^+ \p_w x _+(w)  
\eea
Here, we have defined the transformations $\delta _{ ss}$ on
the fields $x_+$ and $\psi _+$, which act as standard supersymmetry
transformations on the matter fields. Using this supersymmetry transformation
rule, we may compute the various ingredients needed in evaluating
the transformations $\delta _\xi \Y$ as follows. First the transformation
of the supercurrent itself is needed in $\delta _\xi \Y_1$ and is given
by
\bea
\delta _\xi \left \< {1 \over 2 \pi} \int \! \chi S ~S(w) \right \> _{(c)}
& \equiv & \delta _{ ss} S (w) =  \xi ^+ T(w)  
\eea
Since the double supercurrent insertion occurs in a 
connected correlator, the stress tensor insertion above
will also figure only in a connected correlator, and no self-contractions
of $T$ shall occur. One also needs the $\delta _{ss}$ transformations
of the vertex operator components. They are given by
\bea
\delta _{ss} \V ^{(0)} (w) 
& = & 
- dw \p _w \left ( \xi ^+ (w) \e ^\mu \psi ^\mu _+(w) e^{i k \cdot x_+(w)} \right )
\no \\
\delta _{ss} \V ^{(1)} (w) 
& = & 
\half  \xi ^+ (w) \chiw (w) { d \bar w \over dw} \V ^{(0)} (w)
\no \\
\delta _{ss} \V ^{(2)} (w) 
& = & 0
\eea

\subsubsection{Variations of $\Y$}

We are now ready to evaluate all the variations. First, notice
that the $\delta _{ss} S(w)$ term generated in $\delta _\xi \Y_1$
is proportional to the stress tensor insertion and is readily 
cancelled by $\delta _\xi \Y_2$. Therefore, it is useful
to consider right away the following combinations,
\bea
\label{yonetwo}
\delta _\xi \Y_1 + \delta _\xi \Y_2 
=
 {1 \over 2 \pi } \sum _{i=1} ^N  \left \< Q(p_I) \,  \int \!  \chi S ~
 \left (\delta _{ss} \V^{(0)} _i  \right )  
\prod _{j  \not= i}^N \V_j ^{(0)}  \right \> _{(c)}
\eea
The subscript $(c)$ on the correlator  now becomes immaterial
since no self-contractions of $S$ can occur, and 
the subscript will be dropped altogether.

\medskip

The contribution of $\delta _\xi \Y_3$ is evaluated along similar arguments,
and we find,
\bea
\label{ythree}
\delta _\xi \Y_3 & = &
\sum _{i=1} ^N \left \< Q(p_I) \,
\left ( \delta _{ss} \V ^{(1)} _i  \right ) ~ 
\prod _{j \not= i}^N \V_j ^{(0)}  \right \> 
+ \sum _{i \not= j} \left \< Q(p_I) \,  \left ( \delta _{ss} \V_i ^{(0)}  \right )
\V ^{(1)} _j \prod _{l \not= i,j}^N \V_l ^{(0)}  \right \> 
\no \\ && 
+ {1 \over  2 \pi } \sum _{i=1} ^N \left \< Q(p_I) \, \int \!  \chi S ~ 
\left ( \delta _\xi \V ^{(1)} _i   \right )  ~ 
\prod _{j \not= i}^N \V_j ^{(0)} \right \> 
\eea
The $\delta _\xi$  variation of $\V^{(1)}_i$ in the last term of (\ref{ythree})
and the $\delta _{ss}$ variation of $\V^{(0)}_i$ of (\ref{yonetwo})
partially combine into an exact differential, as follows,
\bea
\label{descent}
\delta _\xi \V^{(1)} (w) + \delta _{ss} \V^{(0)}  (w)
& = &
- d_w \left ( \xi ^+ (w) \e ^\mu \psi ^\mu _+ (w) e^{i k \cdot x_+(w)} \right )
\no \\ &&
+ d\bar w \xi ^+(w)  \p_{\bar w} 
\left ( \e ^\mu \psi ^\mu _+(w) e^{i k \cdot x_+(w)} \right )
\eea
The insertion of the last term above into the last term 
of (\ref{ythree}) is given by
\bea
{1 \over  2 \pi } \sum _{i=1} ^N \left \< Q(p_I) \, \int \! \chi S ~ 
d\bar {z_i} \xi ^+(z_i)  \p_{\bar z_i} 
\left ( \e ^\mu \psi ^\mu _+(z_i) e^{i k_i \cdot x_+(z_i)} \right )  ~ 
\prod _{j \not= i}^N \V_j ^{(0)} \right \> 
\eea
The presence of the $\p _{\bar z_i}$ derivative guarantees
that its argument will only give non-vanishing contractions
with the supercurrent (as the points $z_j$ are all distinct from $z_i$
for $j \not= i$). These contractions may be computed,
\bea
\< S(w) \p_{\bar z_i} x_+ (z_i) \>_* 
& = & - \pi \delta (w, z_i) \psi _+ ^\mu (z_i)
\no \\
\< S(w) \p_{\bar z_i} \psi_+ (z_i) \>_* 
& = & - \pi \delta (w, z_i) \p_{z_i} x_+ ^\mu (z_i)
\eea
and as a result,
\bea
\< S(w) \p _{\bar z_i} 
\left ( \e ^\mu \psi ^\mu _+(z_i) e^{i k_i \cdot x_+(z_i)} \right ) \>_*
=
- \pi \delta (w,z_i) \V ^{(0)} _i (z_i)
\eea
Combining all, we have 
\bea
\label{y123}
\sum _{a=1}^3 \delta _\xi \Y_a  
& = &
\sum _{i=1} ^N \left \< Q(p_I) \,
\left ( \delta _{ss} \V ^{(1)} _i  \right ) ~ 
\prod _{j \not= i}^N \V_j ^{(0)}  \right \> 
+ \half \sum _{i =1}^N \xi ^+ \chiz (z_i) \left \< Q(p_I) \, 
~ \prod _{j =1}^N \V_j ^{(0)}  \right \> 
\no \\ && 
+ \sum _{i \not= j} \left \< Q(p_I) \,  \left ( \delta _{ss} \V_i ^{(0)}  \right )
\V ^{(1)} _j \prod _{l \not= i,j}^N \V_l ^{(0)}  \right \> 
+  \sum _{i=1} ^N d_i \R _i ^{\xi \, (1)} [\delta]  
\eea
where the exact differential term is given by
\bea 
\R _i ^{\xi \, (1)} [\delta]   = {1 \over 2 \pi}
\left \< Q(p_I) \, \int \! \chi S ~ \e _i ^\mu \xi ^+(z_i) \psi _+ ^\mu (z_i) 
e^{i k_i \cdot x_+(z_i)} \prod _{j\not= i} \V ^{(0)} _i  \right \>
\eea
The next step is to notice that the first and last terms in (\ref{y123})
cancel $\delta _\xi \Y_5$ completely. Thus, we are left with
\bea
\sum _{a=1}^5 \delta _\xi \Y_a
=
\sum _{i \not= j} \left \< Q(p_I) \,  \left ( \delta _{ss} \V_i ^{(0)}  
+ \delta _\xi \V^{(1)} _i  \right )
\V ^{(1)} _j \prod _{l \not= i,j}^N \V_l ^{(0)}  \right \> 
+  \sum _{i=1} ^N d_i \R _i ^{\xi \, (1)}[\delta]  
\eea
We use again (\ref{descent}) to combine the  two variations in the 
parentheses into an exact differential and a remainder, which involves the 
operator 
$\p_{\bar z_i} \left ( \e ^\mu _i \psi ^\mu _+ (z_i) e^{i k _i \cdot x_+ (z_i) } \right ) $,
whose correlators vanish because of the $\p _{\bar z_i}$
operator acting on a holomorphic expression for separated points $z_j$.
We are left with
\bea
\label{y123a}
\sum _{a=1}^5 \delta _\xi \Y_a
=
\sum _{i=1} ^N d_i \left (\R _i ^{\xi \, (1)} [\delta]   + \R _i ^{\xi \, (2)} [\delta]   \right )
\eea
where the new exact differential term is given by
\bea
\R _i ^{\xi \, (2)} [\delta]   = - \sum _{j \not= i} \left  \< Q(p_I) \,
\xi ^+ (z_i) ~ \e _i ^\mu \psi ^\mu _+ (z_i) e^{i k_i \cdot x_+(z_i)}
\V ^{(1)} _j  \prod _{l \not= i,j} \V ^{(0)} _l
\right \>
\eea

\subsubsection{Structure of the Exact Differential Terms}

Using this last result, we have 
\bea
\delta _\xi \B [\delta] ^{(c)} = 
d\mu _0 [\delta] \sum _{a=1}^5  \delta _\xi \Y_a
=
  \sum _{i=1} ^N d_i \R _i ^\xi [\delta]  
\eea
Here, $\R _i ^\xi [\delta]    = \R _i ^{\xi \, (1)} [\delta]  + \R _i ^{\xi \, (2)}[\delta] $  
is  given succinctly  by the following expression,
\bea 
\label{Rxi}
\R _i ^\xi [\delta]  
= 
 \bigg \<
Q(p_I) \, \O_i ~
\bigg ( {1 \over 2 \pi } \int \! \chi S ~  \prod _{l \not= i} \V ^{(0)} _l
 +
\sum _{j\not= i} \V^{(1)} _j \prod _{l \not= i,j} \V ^{(0)} _l  \bigg ) \bigg \>
\eea
where we have defined the operator
\bea
\O _i \equiv - d\mu _0 [\delta]  ~ \xi ^+ (z_i) ~ \e ^\mu _i \psi _+ ^\mu (z_i) ~
e^{i k_i \cdot x_+(z_i)}
\eea
This form has the following properties
\begin{itemize}
\item $\R _i ^\xi [\delta]  $ is a form of weight $(0,0)$ in $z_i$;
\item $\R _i ^\xi [\delta]  $ is a form of weight $(1,0)\oplus (0,1)$ in 
each $z_j$, with $j\not= i$;
\item $\R _i ^\xi [\delta]  $ is closed in each $z_j$ for $j\not= i$, i.e. 
$d_j\R _i ^\xi [\delta]  =0$.
\end{itemize}
Only the last assertion requires a proof.
We calculate $d_j \R _i ^\xi[\delta]   $ for $j \not= i$.

\medskip

The first term in (\ref{Rxi}) is a form of weight $(1,0)$ in $z_j$, 
and thus only the $d\bar z_j \p _{\bar z_j}$ part of the differential
$d_j$ acts on this term. The only contributions come from the 
further contractions of this derivative with the supercurrent.

\medskip

The second term in (\ref{Rxi}) is a sum of terms of two types. 
The first type consists of those terms involving 
the form $\V ^{(1)} _l (z_l)$ with $l \not= j$; these are forms 
of weight $(1,0)$ in $z_j$, and thus only the $d\bar z_j \p _{\bar z_j}$ 
part of the differential $d_j$ acts. Since, for distinct points $z_l$, 
the argument is holomorphic, this type of terms automatically contributes 0.

\medskip

The second type consists of those terms involving the form
$\V ^{(1)} _j (z_j)$; this form is of weight $(0,1)$ in $z_j$
and thus only the $d z_j \p _{z_j}$ part of the differential $d_j$ acts.
Putting all together, we have
\bea 
d_j \R _i ^\xi [\delta]  
= 
 \bigg \<
Q(p_I) \, \O_i ~
\bigg ( {1 \over 2 \pi } \int  \chi S  \, d\bar z_j \p _{\bar z_j} \V ^{(0)} _j
 +
dz_j \p_{z_j}  \V^{(1)} _j \bigg ) \prod _{l \not= i,j} \V ^{(0)} _l   \bigg \>
\eea
Using the following contraction,
\bea
\left \<  {1 \over 2 \pi } \int \! \chi S ~  d\bar z_j \p _{\bar z_j} \V ^{(0)} _j \right \>_*
= 
- dz_j \p_{z_j} \left ( \V ^{(1)} _j   \right )
\eea
Adding both contributions, we find that
\bea
d_j \R _i ^\xi [\delta]    =0 \hskip 1in j\not= i
\eea
which completes the proof.

\newpage

\section{Slice-Independence of the Full Amplitudes}
\setcounter{equation}{0}

In this section, we use the closedness of the chiral amplitudes and 
exactness of their variations induced by gauge slice changes
proven in the preceding two sections to establish the
gauge slice independence of the GSO projected, vertex integrated,
superstring scattering amplitudes.

\medskip

The chiral blocks $\B [\delta]$ have non-trivial monodromy as each insertion 
point $z_i$ is transported around a $B$-cycle 
(but no monodromy around an $A$-cycle) on the worldsheet. 
This monodromy is due to the monodromy of the effective bosonic propagator
$\< x_+ (z) x_+ (w) \> = -\ln\,E(z,w)$,
\bea
\label{monodromyprop}
- \ln E(z+B_K,w)= - \ln E(z,w)
+ \pi i\Omega_{KK} + 2\pi i\int_w^z\omega_K 
\eea 
The fermionic propagator $S_\delta(z,w)$ has no monodromy (for even spin structures). Carrying out the contractions in $\B [\delta]$, we find that 
$\B [\delta]$ has the monodromy indicated in (\ref{monodromy}) 
(see \cite{dp89}, eq. (5.46)). Formally, this result is readily seen if we 
note that the internal and external momenta-dependence of
$\B [\delta]$ comes from the following 
insertion (c.f. (\ref{block1}))
\bea
Q(p_I)W(z,\theta;\epsilon,k)
=\exp \left ( ip_I^\mu \oint_{B_I} dz \p_z x_+^\mu
+(ik^\mu+\epsilon^{\mu}\theta\p_z) x_+^\mu +
(ik^\mu\theta+\epsilon^\mu)\psi_+^\mu \right )
\no
\eea
Here we have set $(z_1,\theta_1)=(z,\theta)$ and suppressed
the explicit dependence on the other variables, such as $\chi$ 
and $\hat \mu$ for notational simplicity. Transporting $z$ along 
a $B_K$ cycle results in a shift of the exponential by
\bea
(ik^\mu+\epsilon^{\mu}\theta\,\p_z)\oint_{B_K}dz\,\p_zx_+^\mu
=
ik^\mu\,\oint_{B_K}dz\,\p_zx_+^\mu
\eea
which is exactly the same as a shift of $p_I$ by $k^\mu$.
Applying the same arguments to the variations of $\B [\delta] $ 
obtained in section \S 3, we readily see that the forms 
$\R_i ^v [\delta ] $ and  $\R _i ^\xi  [\delta ]  $ induced by changes of 
$\hat\mu$ Beltrami differential and by gauge slice $\chi_\alpha$ 
have the same monodromy (\ref{monodromy}) as $\B [\delta]$.

\medskip

We can give now the proof of gauge slice independence. Consider
first the $\chi$-independence.
In the preceding sections, we have proven the following equations,
\bea
\delta _\xi {\cal B}[\delta] (z; \e,k,p_I) 
& = & 
\sum _{i = 1} ^N d_i \R _i ^\xi [\delta] (z;\e, k,p_I)
\no \\ 
\delta _{\bar \xi} {\cal B}[\delta] (z; \e, k,p_I) 
& = & 
0
\no \\ 
d_i {\cal B}[\delta] (z; \e, k,p_I) & = & 0
\no \\ 
d_j \R _i ^\xi [\delta] (z; \e, k,p_I) & = & 0 \hskip 1in j \not= i
\eea
Here, $d_i$ denotes the total differential in the variable $z_i$. 
Furthermore, $\xi$ denotes the $\xi ^+$ component of the change 
of spinor fields and  $\bar \xi$ denotes its complex conjugate.
Now the GSO projected,
vertex integrated, superstring scattering amplitudes is defined by
\bea
 \sum _{\delta, \bar \delta} \C _{\delta, \bar \delta}
\int dp_I
\int _{\Sigma^N}  {\cal B}[\delta] (z; \e,k,p_I)
\wedge \bar {\cal B}[\bar \delta]' (\bar z; \bar \e, k,p_I)
\eea
Here, the left and right moving strings do not necessarily have to be the 
same (although they are required to have the same monodromy), 
a fact that is indicated by a prime on $\bar \B$.
The coefficients $C_{\delta, \bar \delta}$ govern the GSO 
projection prescription and are independent of the moduli $\Omega$
and of the insertion points $z_1 , \cdots , z_N$.
Under a change $\delta_\xi\chiz(z)=-2\p_{\bar z}\xi^+(z)$
of gauge slice, the contribution to the above amplitude 
from each pair of spin structures $\delta,\bar\delta$ changes to
\bea
&&
\int dp_I
\int _{\Sigma^N}  \bigg({\cal B}[\delta] 
+\sum_{i=1}^Nd_i{\cal R}_i ^\xi [\delta ] \bigg)
\wedge \bigg(\bar {\cal B}[\bar \delta]' 
+\sum_{j=1}^Nd_j\bar{\cal R}_j ^\xi [\delta ] '\bigg)
\nonumber\\
&&
\quad
=
\int dp_I
\int _{\Sigma^N} \bigg({\cal B}[\delta] \wedge \bar {\cal B}[\bar \delta]'
+
\sum_{i=1}^Nd_i \left ({\cal R}_i ^\xi [\delta] \wedge \bar {\cal B}[\bar \delta]' \right )
+\sum_{j=1}^Nd_j \left ( (-)^N{\cal B}[\delta] \wedge {\cal R}_j ^\xi [\delta] ' \right )
\nonumber\\
&&
\qquad\qquad\qquad\qquad
+
\sum_{i,j=1}^N
d_i \left ({\cal R}_i ^\xi [\delta] \wedge 
d_j \bar {\cal R}_j ^\xi [\bar \delta] '\right )\bigg)
\eea
where we have used the fact that ${\cal B}[\delta]$, 
$\bar{\cal B}[\bar\delta]'$ are closed 1-forms in all variables,
while ${\cal R}_i ^\xi [\delta] $, $\bar{\cal R}^{\bar\xi \, '} _ i$
are closed 1-forms in all variables except $z_i$. 
Isolating the integration over $z_i$, we note that the object
${\cal R}_i ^\xi [\delta] \wedge \bar{\cal B}[\bar\delta]'$ is a 1-form in $z_i$ with the monodromy (\ref{monodromy}). Its integral must then vanish by the Riemann identity, upon integration over the
internal loop momenta $p_I$. More precisely,
the surface $\Sigma$ can be cut apart along its homology cycles
into a simply-connected region with boundary $\prod_{K=1}^2A_KB_KA_K^{-1}B_K^{-1}$. Applying Stokes' theorem to the integral of 
$d_i({\cal R}_i ^\xi [\delta] \wedge \bar{\cal B}[\bar\delta]')$ over this region, we obtain
\bea
\label{riemannidentity}
\int_\Sigma
d_i \left ( {\cal R}_i ^\xi  [\delta] \wedge \bar{\cal B}[\bar\delta]' \right )
&=&
\sum_{K=1}^2\oint_{B_K}\bigg(({\cal R}_i ^\xi [\delta] \wedge 
\bar{\cal B}[\bar\delta]')(z_i+A_K;p_I)-
({\cal R}_i ^\xi [\delta ] \wedge \bar{\cal B}[\bar\delta]')(z_i;p_I)\bigg)
\nonumber\\
&- &
\sum_{K=1}^2 \oint_{A_K} \bigg( ({\cal R}_i ^\xi [\delta] \wedge \bar{\cal B}[\bar\delta] ')(z_i+B_K;p_I)-
({\cal R}_i ^\xi [\delta] \wedge \bar{\cal B}[\bar\delta] ')(z_i;p_I) \bigg)
\no
\eea
Here we have indicated the dependence on $p_I$ because it plays an important role. In view of the monodromy (\ref{monodromy}) for ${\cal R}_i ^\xi [\delta] \wedge \bar{\cal B}[\bar\delta]'$, the above expression reduces to
\bea
-\sum_{K=1}^2 \oint_{A_K} \bigg( ({\cal R}_i ^\xi [\delta] \wedge 
\bar{\cal B}[\bar\delta] ')(z;p_I+\delta_{IK}k_i)-
({\cal R}_i ^\xi [\delta] \wedge \bar{\cal B}[\bar\delta] ')(z;p_I)\bigg)
\eea
For fixed $p_I$, this may not vanish, but its integral over $p_I$ does
vanish by translation invariance of the measure $dp_I$ and the 
integration region of $p_I$,
\bea
\int dp_I
\sum_{K=1}^2\oint_{A_K}\bigg(({\cal R}_i ^\xi  \wedge 
\bar{\cal B}[\bar\delta]')(z;p_I^{\mu}+\delta_{IK}k_i^\mu)-
({\cal R}_i ^\xi \wedge \bar{\cal B}[\bar\delta]')(z;p_I^\mu)\bigg)=0,
\eea
Thus the contribution of the term 
$d_i({\cal R}_i ^\xi [\delta] \wedge \bar{\cal B}[\bar\delta]')$ is $0$. The contributions of the other exact differentials in (\ref{riemannidentity}) vanish by the same argument, establishing the desired invariance of the integrated superstring
amplitudes under changes of $\chi$ slice.
The situation for changes of $\hat\mu$ slices is similar, starting from the basic properties
\bea
\delta _v {\cal B}[\delta] (z; \e,k,p_I) 
& = & 
\sum _{i=1} ^N d_i \R_i ^v [\delta] (z; \e, k,p_I)
\no \\
\delta _{\bar v} {\cal B}[\delta] (z; \e,k,p_I) 
& = & 
0
\no \\ 
d_j \R _i ^v [\delta] (z; \e,k,p_I) & = & 0 \hskip 1in j \not= i
\eea
The proof of gauge invariance under changes of both $\chi_\alpha$ and 
$\hat\mu$ slices is now complete.

\end{document}